\documentclass[letterpaper]{article}

\usepackage{epsfig}

\begin{document}

\title{Neutron Diffraction and Optics of a Noncentrosymmetric
Crystal. New Feasibility of a Search for Neutron EDM}

\author{V.V. Fedorov, V.V. Voronin\footnote{Petersburg Nuclear Physics Institute, 188300, Gatchina,
 St.Petersburg, Russia, E-mail: vvv@mail.pnpi.spb.ru. This work is a basis for Dr. of Sci. thesis of  V.V.Voronin} }

\maketitle

\begin{abstract}
Recently strong electric fields (up to $10^9$ V/cm) have been
discovered, which affect the neutrons moving in noncentrosymmetric
crystals. Such fields allow  new polarization phenomena
in neutron diffraction and optics and provide,
for instance, a new feasibility of a search for the neutron electric dipole
moment (EDM).  A strong interplanar electric field of the crystal
and a sufficiently long time for the neutron passage through the crystal
for Bragg angle close to $\pi/2$ in the case of Laue diffraction
make it possible to reach the sensitivity achieved with the most sensitive now
magnetic resonance method using ultra cold neutrons (UCN method).

A series of  experiments was carried out in a few last years on study of the dynamical diffraction of
polarized neutrons in thick (1--10 cm) quartz crystals, using the forward
diffraction beam and Bragg angles close to 90$^\circ$. As well new neutron optics phenomena were investigated.
The following
effects was first observed:

$\bullet$\ the effect of an essential time delay of a diffracting neutron
inside the crystal for Bragg angles close to $\pi/2$;

 $\bullet$\ the phenomenon of a neutron beam depolarization 
 in time of Laue diffraction by noncentrosymmetric $\alpha$-quartz
crystal;

$\bullet$\ the effect of a neutron-optical spin rotation for neutrons
 moving through a noncentrosymmetric crystal with the energies  and directions far
 from the Bragg ones;
 
$\bullet$\ the controlled spin rotation effect for the Bragg reflected (from a smal part of crystal near its exit face) neutrons in a slightly deformed quartz crystal.
 
The feasibility of  experiment on a search for  neutron EDM using Laue diffraction in crystals without a center of symmetry was tested at the reactors: WWR-M in Gatchina and HFR in Grenoble. It was shown that the sensitivity can reach $(3 - 6)\cdot 10^{-25}~e\cdot{\rm cm}$
per day for the available quartz crystal and cold neutron beam flux.
 
\end{abstract}
 
\section{Introduction}

 The origin of CP-symmetry  violation (where C is a charge conjugation and P
 is a spatial inversion) is  of a great interest since its discovery in the
 decay of neutral K-mesons about
 40 years ago.  CP violation leads in turn to the violation of the time
 reversal symmetry (T) through the CPT invariance (CPT-theorem).
 Existence of nonzero neutron EDM requires violation of both P and T
 invariance. Different theories of CP violation give widely varying
 predictions for a neutron EDM (see for instance \cite{Khriplovich,Bunakov}). 
 Therefore, the new experimental limits on the EDM
 value would be of great importance for understanding
the nature of the CP violation as well as of the Universe baryon asymmetry, which are beyond the Standard Model.

 The most precision method of the EDM measurement now is the magnetic
 resonance one \cite{pnpiedm,illedm,edmlast}\footnote{Last
experimental result for a neutron EDM obtained by this method is
$d_n\leq 6.3\cdot 10^{-26}$ e$\cdot$cm at the 90\% confidence
level \cite{edmlast}.} using
 the ultracold neutrons (UCN method).

Here we shall discuss another possible way for the neutron EDM search,
using  Laue diffraction of the  cold neutrons in the noncentrosymmetric crystal \cite{NIMB_FedVor}.
Earlier \cite{dfield} we have shown that the neutron,
moving through a noncentrosymmetric crystal, may be influenced
with a strong interplanar electric field. Its value depends on 
direction and  value of a neutron wave vector, reaching a maximum
(up to $10^9$ V/cm), when the Bragg condition for some system
of crystallographic planes is satisfied. These fields result in
new polarization phenomena observable in  neutron diffraction and
optics. In particular, the Schwinger interaction of the neutron
magnetic moment with such fields leads to spin dependence of the
neutron Pendell\"osung picture \cite{dfield}. The Pendell\"osung phase
shift accompanying the spin flip was  measured first in ref. \cite{dfield},
and so the interplanar electric field was determined in the experiment on
dynamical Laue diffraction of polarized neutrons for the (110)-plane
of the $\alpha$-quartz crystal.
Experimental value of the field  was obtained to
be equal $E_{(110)}=(2.10\pm 0.12)\times 10^8$~V/cm and
had coincided with the theoretical one.

The interplanar electric fields are more than four
orders of magnitude higher than those used in the UCN
methods \cite{pnpiedm,illedm,edmlast} of the neutron EDM search. So
it was a natural idea \cite{dedm,polart,Forte} (arisen anew
after Shull and Nathans \cite{Shull})
to use these crystal fields for searching the neutron EDM
(see also \cite{Barysh,GolPendl}).
However, the value of the crystal field turned out to be still
insufficient to reach the sensitivity of the UCN method,
but it was shown \cite{dedm} that use of Laue diffraction
for EDM measurements with the Bragg angles close
to $90^\circ$  would essentially increase the time, the neutron
spends in crystal under the strong electric field.

Two variants of the method, using the Laue diffraction of polarized neutrons in
the crystals without a centre of symmetry, were proposed  for this
purpose. One of them is the double crystal variant \cite{dedm}
based on a spin dependence  of the Pendell\"osung picture phase and
the second is the polarization method \cite{polart} using the
depolarization effect for neutrons diffracting in the
noncentrosymmetric crystals.
It was shown \cite{dedm,polart} that  the sensitivity
of the method to measure the neutron EDM may be increased more than by an order of magnitude
by a choice of Bragg angles close to $\pi/2$, so it may exceed
(with the higher luminosity taken into account) the sensitivity of the UCN
method \cite{pnpiedm,illedm,edmlast}.
That is possible for Laue diffraction scheme only.
But only the experimental study of these effects  can answer the question on
actual sensitivity of the method to measure the neutron EDM.

A hypothetical idea to use the interplanar electric fields (if they did exist)
for the neutron EDM search was discussed in the review \cite{GolPendl}.
But such crystal properties were not known at that
time.

The importance  to consider the crystal noncentrosymmetricity
was previously pointed out in ref. \cite{3}.
Authors \cite{3} were 
the first who paid attention to an existence of
the interference between the electric and nuclear structure
amplitudes for neutrons diffracted by noncentrosymmetric
crystal and proposed to study  the Schwinger interaction using such crystals. 

In the work \cite{Forte} the effect of neutron spin
rotation due to such interference has been discussed.
Similar, but more detailed theory of neutron optical activity and
dichroism for diffraction in noncentrosymmetric crystals has
been developed in ref. \cite{Barysh}.

It has been shown \cite{Forte} that the spin
rotation  effect in a non-absorbing crystal can take place only for
Bragg scheme of diffraction.  The deviation from the Bragg condition by
about the Bragg width  is
necessary in this case to observe the effect for the transmitted beam. That
reduces the effect, because it is proportional to $1/\sqrt{1+w^2}$,
where $w$ is the parameter of angular deviation (measured in the
units of Bragg halfwidth) from the Bragg direction, see~\cite{dedm}.
The nuclear absorption is necessary to have a spin rotation
effect in the case of Laue diffraction \cite{Forte,Barysh}.

The possibility of a search for neutron EDM by measuring a spin
rotation angle was analyzed for the Bragg diffraction
scheme \cite{Forte} and for the Laue scheme \cite{Barysh}.

We have shown \cite{dfield} that for some system of crystallographic
planes in noncentrosymmetric crystals the positions of the
electric potential maxima can be shifted relative to those of a nuclear
potential. Hence, the diffracted neutrons
will move in the crystal under a strong ($10^8 - 10^9$~V/cm)
interplanar electric field (because of the neutron concentration  on
the nuclear potential maxima or between them).
Such a concept turned out to be
very fruitful for further consideration and understanding the
different phenomena, concerning the neutron diffraction and the
optics in noncentrosymmetric crystals \cite{dfield,dedm}. For
example it allowed to predict and to give a simple description of
such new effects as the spin dependence of the pendulum phase,
the depolarization of the diffracting neutron beams, the independence
of the effects due to the Schwinger interaction on the neutron
wavelength and Bragg angle for given crystallo\-graphic planes
etc. \cite{polart,sdprepr,fedvor}.

We should note also, that the Bragg reflection of neutrons from
the neutron-absorbing centrosymmetric crystal of CdS
was used  earlier  \cite{Shull} for EDM search. But the
sensitivity  of this method is much lower than that of
UCN-method, because the depth of neutron penetration into the
crystal, which  determines a time the neutron stay in crystal,
was very small (about 7$\cdot 10^{-2}$~cm).
Now a new variant of the method is proposed and developed,
using  the multiple reflections from the silicon centrosymmetric
crystal \cite{Rauch1,Domb}.

In the work \cite{ForteZeyen} it was reported that the effect of the neutron spin rotation has been
observed  due to the spin-orbit (Schwinger) interaction, using the Bragg scheme of the
diffraction in the noncentrosymmetric crystal with a small deviation
of a neutron momentum direction (by about a few Bragg width) from the Bragg
one, because  the effect disappears for the exact Bragg direction in this case.
However,
the experimental value
of the spin rotation angle \cite{ForteZeyen} turned out to be
a few times less than the theoretical one. Authors were in difficulty to explain the  origin of such discrepancy, but it 
was very likely due to imperfection of the used crystal.

Here we consider neutron-optic effects for neutrons, moving
through a noncentrosymmetric crystal with the energies and  the
directions far from the Bragg ones, when the deviation from the exact Bragg
condition reaches $(10^3 - 10^5)$ Bragg widths.

The theoretical estimations have shown \cite{PbTiO3}, that for the
polar noncentrosymmetric crystal ($PbTiO_3$, for instance) the
value of a resultant electric field, acting on a neutron,
can reach $\approx 2\times 10^6$ V/cm for wide range (about four orders
exceeding the Bragg width) of the neutron directions and wavelengths.
Such a field is a result of superposition of the fields from a few
different crystallographic planes. Therefore, the spin rotation effects
may be not too small in neutron optics in comparison with the
diffraction case.

The observation of  such effects may be of interest for
development of new methods for searching  the neutron EDM. It
also may be useful for experimental searching the T-odd part of
the nuclear interaction, using neutrons with energy close to the
P-resonance one \cite{bar}, because it is hardly possible to
observe dynamical diffraction in a large crystal for neutrons with
the energy $\sim 1$~eV because of too high requirements to a crystal
quality.

The phenomenon can be used also for measurements of the
interplanar electric fields affecting the neutrons in the
crystals without centre of
symmetry. A new specific spin neutronography arises in this case
for crystallographic planes with nonzero electric fields.

\section{Diffraction in a noncentrosymmetric crystal}

As it follows from the dynamical diffraction theory, a movement
of a neutron through the crystal in a direction close to the
Bragg one for some system of crystallographic planes can be
described by two kinds of Bloch waves $\psi ^ {(1)}$ and
$\psi^{(2)}$ (see for example, \cite {Hirsh}), formed as a result
of neutron interaction with the periodic nuclear potential
\cite{dedm} $V^N_g({\bf{r}})= V^N_0+2v^N_g\cos({\bf{gr}})$, where
${\bf{g}}$ is a reciprocal lattice vector describing the system
of crystallographic planes, $|{\bf{g}}|=2\pi/d$, $d$ is an
interplanar spacing, $V_0^N$ is the average nuclear potential
of the crystal,

\begin{equation}
    \psi^{(1)}=\cos\gamma e^{i\mbox{\boldmath $k$}^{(1)}\mbox{\boldmath $r$}}
+\sin \gamma e^{i\mbox{\boldmath $k$}^{(1)}_g\mbox{\boldmath $r$}}  ,
\label{eq:34}
\end{equation}

\begin{equation}
\psi^{(2)}=
       -\sin\gamma e^{i\mbox{\boldmath $k$}^{(2)}\mbox{\boldmath $r$}}+
   \cos\gamma e^{i\mbox{\boldmath $k$}^{(2)}_g\mbox{\boldmath $r$}},
\label{eq:35}
\end{equation}
where $\mbox{tg}\, 2\gamma=|U_g|/\Delta\equiv 1/w =\gamma_B/\Omega;\;
0 < \gamma < \pi/2$, $U^N_g=2mv_g^N/\hbar^2$, $\mbox{\boldmath $k$}^{(1,2)}_g= \mbox{\boldmath $k$}^{(1,2)}+\mbox{\boldmath $g$}$; $\Delta = (k^2_g - k^2)/2$ . The parameter $w$ is the ratio of angular deviation from the exact Bragg direction $\Omega=\theta-\theta_B$ to the angular Bragg width $\gamma_B$ , therefore $w$ describes  the relative deviation from the exact Bragg condition. 

The intensities of direct and reflected waves in the states (\ref {eq:34}), (\ref {eq:35}) are given by
\[
    \cos^2\gamma =
  \frac{1}{2}[1+ \frac{\Delta_g}{\sqrt{\Delta_g^2+|U_g|^2}}]=
\frac{1}{2} [1+\frac{w}{\sqrt{1+w^2}}].
\]

\begin{figure}[htbp]
	\centering
		\includegraphics[width=0.75\textwidth]{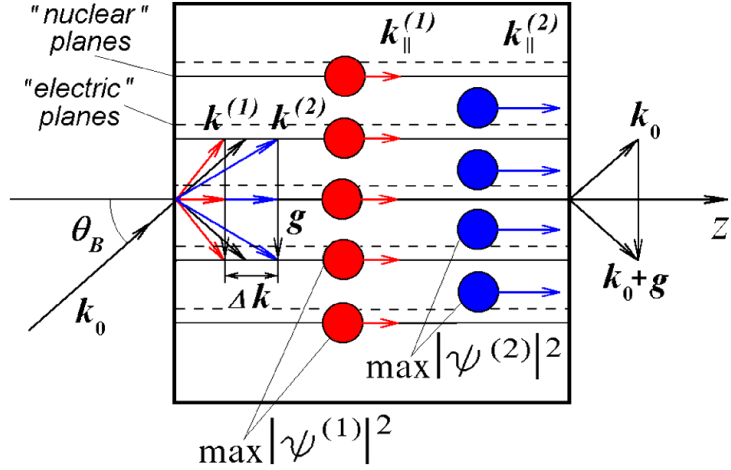}
\caption{Passage neutron through the crystal.}
\label{fig:00}       
\end{figure}

The expressions (\ref {eq:34}), (\ref {eq:35}) describe  two
standing waves (in
the direction ${\bf{g}}$ normal to the crystallographic planes),
 which are moving in the direction
${\bf{k}}_\|^{(1,2)} = {\bf{k}}^{(1,2)} + {\bf{g}}/2 $ along the
planes (see Fig. \ref{fig:00}). 
A small difference of the wave vectors $k^{(1)}$,
$k^{(2)}$ is a result of neutron concentration on "nuclear"
planes (for $\psi^{(1)}$) and between them (for $\psi^{(2)}$),
\begin{equation}
     k^{(1,2)\,2}=K^2 -\Delta \pm \sqrt{\Delta^2 + |U_g|^2}.
\end{equation}
 where 
$K^2=k_0^2+U_0^N \equiv 2m(E+V_0^N)/\hbar^2$, $m$, $E$, $k_0$ are
respectively the mass, energy and wave vector of the incident
neutron.

The values $V_0^N$, $v_g^N$ have an order of $10^{-7}$ eV, so for
thermal and cold neutrons with the energies of
$(10^{-1}-10^{-3})$~eV we can consider that $k^{(1)}\approx
k^{(2)}\approx k$. The propagation velocities of these waves along
the crystallographic planes are\footnote{Here we neglect the
Pendell\"osung oscillations (arising from the interference of waves of
different type) because in our case they are averaged over Bragg angles
for a slightly divergent beam.}
\begin{equation}
v_{\|}^{(1,2)}=\frac{\hbar}{m}|{\bf{k}}^{(1,2)}
+ {\bf{g}}/2|=\frac{\hbar}{m} k^{(1,2)} \cos \theta_B\approx
v \cos \theta_B,
\end{equation}
where $v=\hbar k /m=2\pi\hbar/ (\lambda m)= \pi \hbar / (m d\sin
{\theta_B})$ is the velocity of the incident neutron, $\theta_B$
is the Bragg angle, $\lambda =2\pi/k$ is the wavelength of the
incident neutron ($\lambda = 2d \sin {\theta_B}$). A number of
the dynamical diffraction phenomena (see, for example,
\cite{Hirsh,Kirian,RauchPetr}),
including effects caused by the neutron EDM \cite{dedm,polart},
are  determined not by  a total neutron velocity $v$, but its
component along the crystallographic planes $v_{\|}=v\cos\theta_B$.
In particular, the time the diffracting neutron spends in
crystal, which equal to $\tau_L=L/(v\cos\theta_B)\approx
L/[v(\pi/2-\theta_B)]$, sharply grows for Bragg angles close to
$\pi/2$, where $L$ is the thickness of a crystal. That allows to
increase the sensitivity of the diffraction method to neutron EDM at
least by an order \cite{dedm,polart}. Therefore the values
\footnote{The set-up sensitivity is determined by $\sigma(D)
\propto 1/E\tau \sqrt N$, where $\sigma(D)$ is an absolute error
of EDM measurement, N is a total value of accumulated events.}
$E\tau$ can be of the same order for UCN and diffraction method
(for Bragg angles sufficiently close to $\pi /2$) \cite{dedm,polart}
 despite the fact that the storage time for UCN
($\sim 100$ s \cite{pnpiedm}) is
essentially more than the time of a neutron passage through the
crystal\footnote{The ways of possible improving the UCN-method
are discussed \cite{Golub}.}.

We proposed the polarization method for searching a neutron EDM based
on the predicted  effect of depolarization of the diffracted
neutron beam \cite{polart} in noncentrosymmetric crystal. This method has some advantages (such
as its relative simplicity and less sensitivity to crystal
imperfection) over the method based on the spin dependence of the
neutron Pendell\"osung phase \cite{dedm}.

\begin{figure}[htbp]
	\centering
		\includegraphics[width=0.75\textwidth]{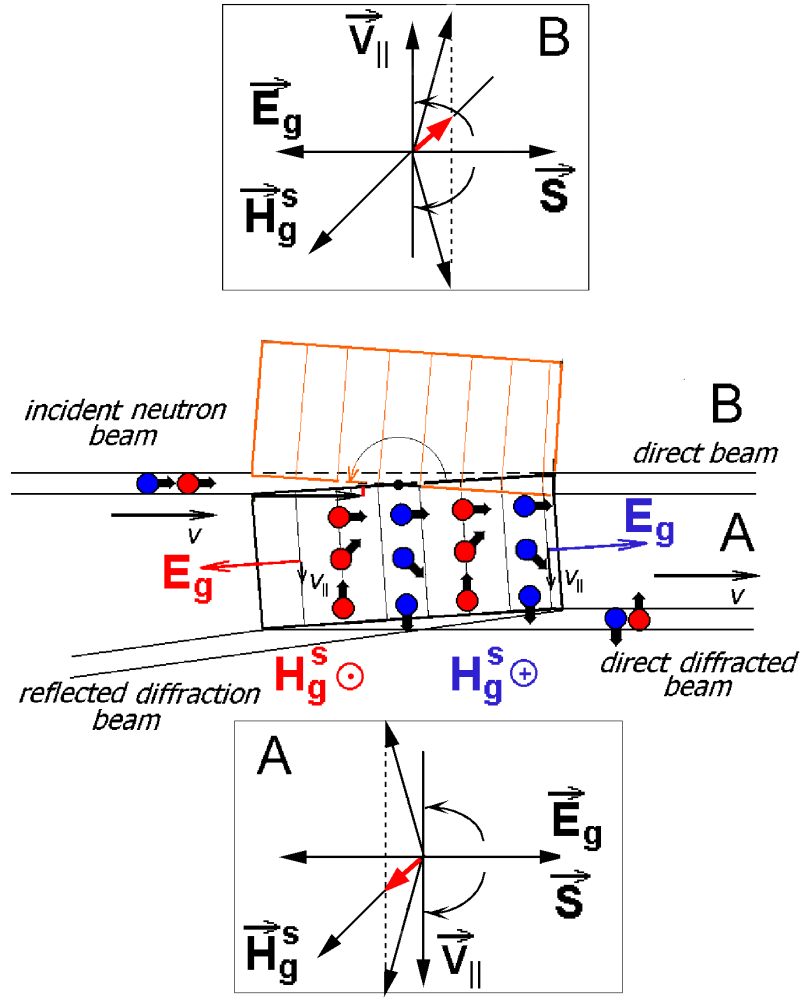}
\caption{
The  neutron
spin for two Bloch waves (shown schematically as the red and blue circles)
will rotate in opposite direction under opposite fields. When the angles of
spin rotation become
 equal to $\pi/2$, the both diffracted neutron beams will be depolarized entirely.
The existence of neutron EDM will result in a polarization along
the Schwinger magnetic field. The sign of this polarization will be different for two crystal position A and B.}
\label{fig:0}       
\end{figure}

Essence of the effect is as follows. In the noncentrosymmetric
crystal the diffracting neutrons in
two Bloch states are moving under opposite electric fields \cite{dfield,dedm,sdprepr} 
 because of the shift of the "electric" planes relative
 to the "nuclear" ones, see Fig. \ref{fig:00}.
  We mean that "nuclear" or "electric" planes are
 determined by the
maximum positions of the corresponding periodic potentials of this plane system 
$$
V^N_g({\bf{r}})= V^N_0+2v^N_g\cos({\bf{gr}}),
$$ 
$$
V^E_g({\bf{r}})= V^E_0+2v^E_g\cos({\bf{gr}}+\phi^E_g),
$$ 
the shift $\phi^E_g$ may be calculated using the crystal structure. 
The electric field of the plane system has a form
$$
{\bf {E(r)}}= - \mbox{grad }V^E_g({\bf{r}})= 2 v^E_g {\bf g}\sin({\bf{gr}}+\phi^E_g).
$$
So the mean electric fields acting on a neutron in the states $\psi^{(1)}$ and $\psi^{(2)}$ will be equal
\begin{equation} 
\langle\psi^{(1)}|{\bf {E(r)}}|\psi^{(1)}\rangle = -\langle\psi^{(2)}|{\bf E(r)}|\psi^{(2)}\rangle   = \frac{{\bf E}_g}{\sqrt{1+w^2}}=\frac{{\bf g}v^E_g\sin\phi^E_g}{\sqrt{1+w^2}}.
\end {equation}
where
\begin{equation} 
{\bf E}_g\equiv  {\bf g}v^E_g\sin\phi^E_g,
\end {equation}
is the maximum electric field acting on a neutron under the exact Bragg condition ($w=0$).

Therefore spins in these
states will rotate in the opposite directions due to Schwinger interaction,
 that in turn will lead to
a decrease of the  neutron beam polarization (see Fig.~\ref{fig:0}).
If an initial spin orientation is
normal to the "Schwinger" magnetic field ${\bf{H}}_g^S=
[{\bf{E}}_g\times{\bf{v}}_{\|}]/c$, then for the case $w<<1$ (exact Bragg condition) the
spin rotation angle in both states will be equal~to
\cite{dedm,sdprepr}
\begin{equation}
\Delta\phi_0^S=\pm\frac{2\mu H_g^S L}{\hbar
v_{\|}}=\pm\mu_n\frac{eE_gL}{m_pc^2},
 \label{DfiS}
\end{equation}
because ${\bf{E}}_g \perp {\bf{v}}_{\|}$ and $H^S_g
=E_g v_{\|}/c$.
Here the signs $\pm$ are related to different states
(1), (2) respectively, $\mu_n=-1.9$ is the neutron magnetic moment in nuclear
magnetons.
As a result the value of neutron beam polarization $P$ will depend
 on $\Delta\phi_0^S$ in the following way\footnote{This result is
 obtained averaging   the Pendell\"{o}sung oscillations
over Bragg angles. The angular period
of this  oscillations in our case is $\sim 10^{-5}$~rad and  the
angular divergence of the neutron beam is $\sim 10^{-2}$~rad.}:
\begin{equation}
      P = P_0  \cos \Delta\phi_0^S=
P_0 \cos\left({\mu_n  e E_g  L \over m_p  c^2}\right),
\end{equation}
$P_0$ is the incident beam polarization (see Fig.~\ref{fig:0}).

The polarization $P$ can be decreased down to zero by a choice
of such a crystal thickness $L_0$ that makes the spin rotation
angles equal to $\pm\pi/2$. Theoretical calculation for
$(110)$-planes of $\alpha$-quartz gives $L_0=3.5$~cm.

The existence of the neutron EDM leads to a slight
polarization $P_{EDM}$ along ${\bf{H}}_g^S$, equal~to
\begin{equation}
P_{EDM}= \frac{4DE_gL_0}{\pi \hbar {v_\|}}=\frac{4D}{\pi\mu}\cdot\frac{c}{v \cos
\theta_B}
\propto \frac{1}{\pi/2-\theta_B},
 \label{Ph}
\end{equation}
because $\cos \theta_B \propto \pi/2-\theta_B$ for $\theta_B
\rightarrow \pi/2$. Here  $D$ is the
neutron EDM. The turn of the crystal by the
angle $2\theta_B$  (see Fig.~\ref{fig:0}) will change the $P_{EDM}$ sign but
will not do that for residual polarization. High precision of the crystal
turning (better than $10^{-5}$ rad) gives the
possibility to exclude the systematic errors and to select the EDM effect.

\begin{figure}[htbp]
	\centering
		\includegraphics[width=0.80\textwidth]{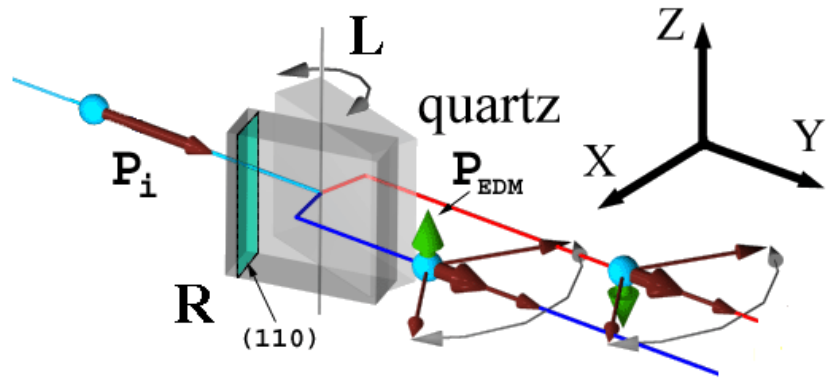}
    \caption{Scheme of the experiment for a neutron EDM search by Laue diffraction.
      The presence of a neutron EDM will lead to a small Y-component of
      the polarization, which will have different signs for the two crystal
      positions R and L.}
    \label{fig:Fig1}
\end{figure}

The principal scheme of the Laue diffraction method to search for the neutron EDM \cite{PhysB2003,Appl_Ph_DEDM} is shown in Fig.~\ref{fig:Fig1}. For two
crystal positions R and L with the same Bragg angle but with 
opposite directions of the electric field, the polarization $P_{\rm EDM}$ will
have opposite signs whereas a residual polarization
will have the same sign for both crystal positions. Therefore we should put the initial neutron spin along the neutron velocity (Y axis) and compare the components of the polarization along the Z axis for two crystal position marked by R and L in Fig.~\ref{fig:Fig1}. 

As it follows from (\ref{DfiS}) the effect due to
the Schwinger interaction does not depend on such neutron
properties as the energy, wavelength and the Bragg angle. It is
determined  by the property of crystal and by the fundamental
constants only. For given crystal it is the same for any Bragg
angles. That gives an
additional way to eliminate a false effect concerned with the Schwinger
interaction  by carrying out the
measurements for two Bragg angles, for example.
The EDM effect (\ref{Ph}) (in contrast to the Schwinger one) depends on
the Bragg angle. It essentially increases
for $\theta_B \rightarrow \pi/2$.

\section{Measurement of time the neutron spends in crystal}
\begin{figure}[htbp]
	\centering
		\includegraphics[width=0.85\textwidth]{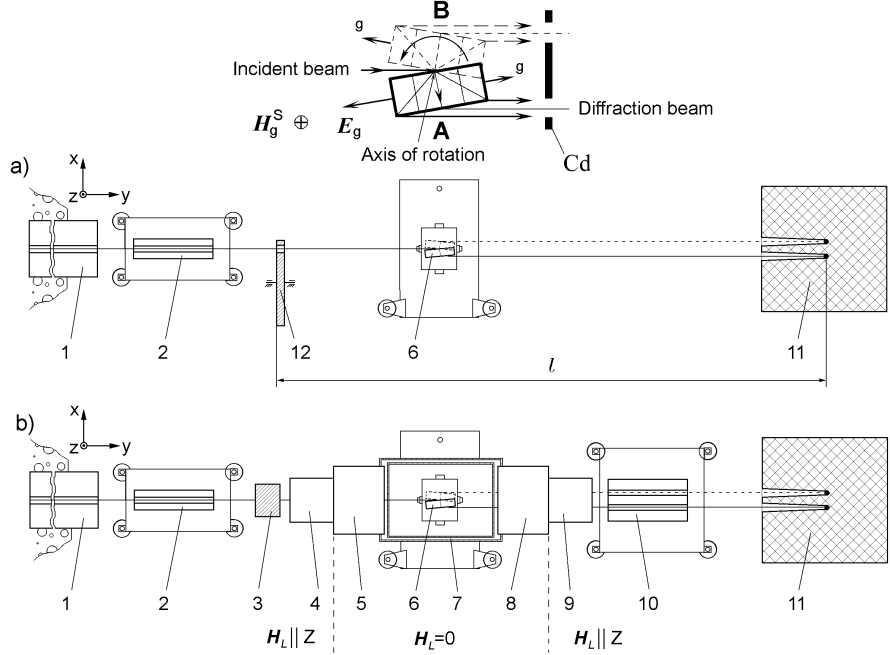}
\caption{Two modifications of the experimental set-up:
{\bf a)} for time-of-flight measurements,
{\bf b)} for measurement of the depolarization effect.
{\bf 1} is a reactor interchannel neutron guide, {\bf 2} is a
multislit polarizing neutron guide, {\bf 3} is the $BeO$ polycrystal filter
(120~mm), {\bf 4,9} are spin-guide coils, {\bf 5,8} are
spin-rotation coils, {\bf 6} is the $\alpha$-quartz single crystal,
(sizes are $14\times 14\times 3.5$ cm$^3$),
{\bf 10} is a double
multislit polarizing neutron guide, {\bf 11} are neutron
detectors, {\bf 12} is a beam chopper. {\bf A} and {\bf B} are
two crystal positions with the same Bragg angles, {\bf g} is the
reciprocal lattice vector for the (110)-plane, $\bf{H_L}$ is the
guiding magnetic field, $l$ is the TOF length.}
\label{fig:1}       
\end{figure}

Two experiments described below on observing the effect of the neutron time delay in the crystal and the depolarization effect were carried out at the WWR-M reactor in PNPI (Gatchina, Russia).
The scheme of experimental setup is shown in Fig.~\ref{fig:1}$^{a)}$   \cite{tfjetpl,Nopt_NIMB}.

The neutron beam formed by neutron guides 1,2 is diffracted by
the noncentrosymmetric $\alpha$-quartz crystal 6
(the reflecting (110) planes  are
normal to the large crystal surfaces) and is registered by the
detectors 11.

\begin{figure}[htbp]
	\centering
		\includegraphics[width=0.75\textwidth]{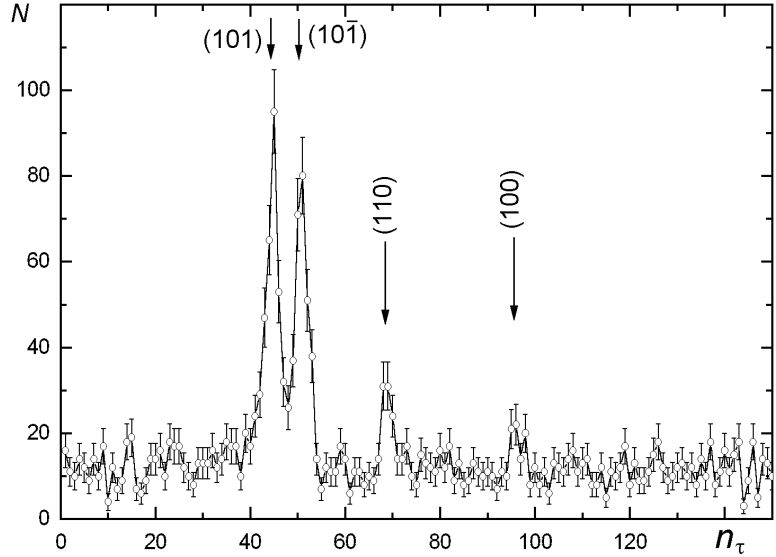}
\caption{TOF spectrum of the forward diffracted neutrons for
Bragg angle~$\theta_{B}=75^\circ$. $n_{\tau}$ is the order
number of the TOF channel. The width of the TOF channel is equal
$\simeq 51.2~\mu$s. N is the number of accumulated events.}
\label{fig:2}       
\end{figure}

All neutrons diffracted by the different crystallographic plane
systems (for which the Bragg conditions are satisfied) give the
contribution to the intensity of the direct diffracted beam.  We used the
time-of-flight (TOF) technique to select the specified reflection. The mechanical
beam chopper 12 was placed before the crystal. The typical TOF
spectrum is shown in Fig.~\ref{fig:2}. The peaks corresponding to neutrons
diffracted by the different crystallographic planes are clearly visible in the
figure.

If the crystal is located between  the beam chopper  and the neutron
detector, the total time of flight of neutron with the
wavelength $ \lambda = 2 d \sin \theta _ B $ will be:
\begin{equation}
 \tau_f=\tau_{\hspace{0.5mm} l}+\tau_L=\frac{l}{v}+\frac{L}{v\cos\theta_B}=
\frac{d\hspace{0.5mm}m}{\hbar\pi}(l\sin{\theta_B}+
L\hspace{0.5mm}\mbox{tg}\hspace{0.5mm}
 \theta_B),
 \label{tfull}
\end{equation}
where $\tau_{\hspace{0.5mm} l}$ is the neutron time of flight for a
distance $l$, $\tau_L$ is the time a neutron spends in the
crystal, $L$ is the thickness of the crystal, $\theta_B$ is the Bragg
angle (for the (110) plane of the $\alpha$-quartz crystal  $d=2.4564$\AA).
As follows from (\ref {tfull}) the time $\tau_L$ of a neutron delay
in the crystal depends on the Bragg angle as
$\mbox{tg}\hspace{0.5mm}\theta_B$, while the
$\tau_{\hspace{0.5mm}l}\propto \sin{\theta_B}$, so $\tau_L$ may
give a very essential contribution to the total TOF of neutrons
$\tau_f$ for $\theta_B$ close to $\pi/2$, because
$\tau_L/\tau_{\hspace {0.5mm}l} \simeq L/[l(\pi/2-\theta_B)]$.

\begin{figure}[htbp]
	\centering
		\includegraphics[width=0.75\textwidth]{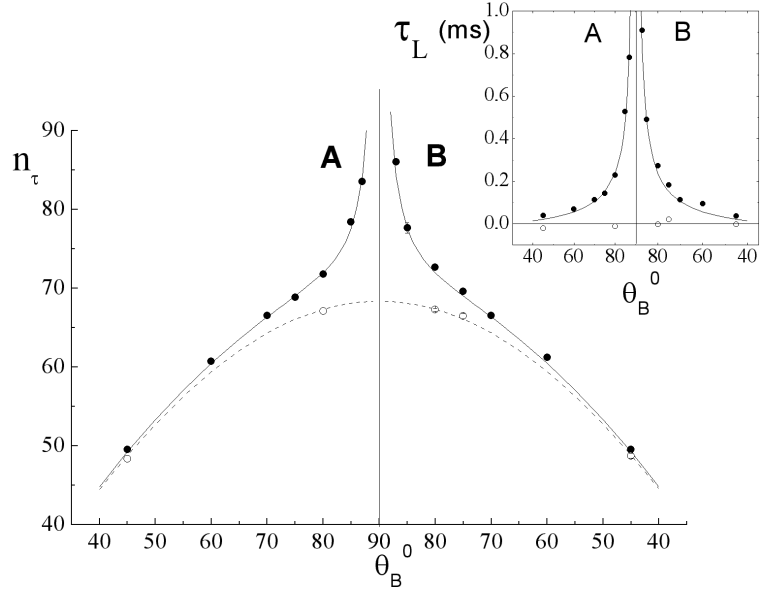}
\caption{Dependence of the neutron TOF on the Bragg angle
for forward diffracted beam.}
\label{fig:3}       
\end{figure}

A dependence on the Bragg angle of the neutron TOF for
forward beam diffracted by the (110)-plane  is given in Fig.~\ref{fig:3}.

A solid curve describes the calculated dependence $\tau_f(\theta_B)$
(see (\ref{tfull})), and the dotted curve does that for
$\tau_{\hspace{0.5mm} l}(\theta_B)$. One can see a good agreement
between experimental (black points) and theoretical dependencies
$\tau_f(\theta_B)$. A control experiment, when the chopper of
a neutron beam  was placed between the crystal and detector,
was also carried
out. In this case the delay of neutron in the crystal does not give
a contribution to the measured value and the position of the
line $\tau_{\hspace{0.5mm} l}(\theta_B)$ for (110)-reflection
should coincide with the dotted curve. That is what we have observed
experimentally (open points). On insertion in Fig.~\ref{fig:3} the
theoretical and experimental dependencies $\tau_L(\theta_B)$
are shown.

So the experiment have proved that the time of neutron delay in
the crystal is not determined by the total neutron velocity $v$, but
its component $v_{\|}$ along the crystallographic plane. In
particular, for $\theta_B=87^\circ$ we find that $\tau_L=(0.90\pm
0.02)$ ms and $v_{\|}=(39\pm 1)$ m/s, while $v=808$ m/s.

\section{Measurement of the depolarization effect}

So we have shown that the polarization of the diffracted beams will be zero for crystal thickness $L=3.5$ cm, if the initial neutron polarization is perpendicular to the Schwinger magnetic field ${\bf{H}}^S_g$, while for initial polarization parallel to ${\bf{H}}^S_g$ will stay completely polarized.  We used a set of different angles between directions of initial neutron polarization and Schwinger magnetic field to study this effect in detail.

\begin{figure}[htbp]
	\centering
		\includegraphics[width=0.80\textwidth]{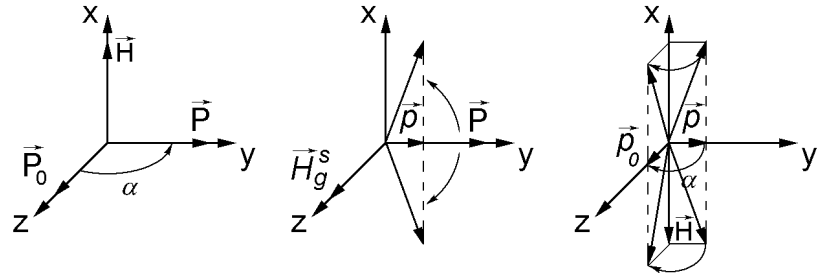}
\caption{The behavior of the diffracted neutron spin for the case
$\alpha=90^\circ$.}
\label{fig:4}       
\end{figure}

The scheme of the experimental setup is shown in Fig.~\ref{fig:1}$^{b)}$, 
see also \cite{Nopt_NIMB,dptfe}. The polarization vector ${\bf P}_0$ of a neutron 
 after passage  through the
polarizing neutron guide 2 and the filter 3 is directed along
${\bf{H}}^S_g$ by  the coil 4, then it  turns round by the angle $\alpha$
by the coil 5. If the crystal does not influence the
spin orientation, the polarization vector will be restored
in the initial direction along the ${\bf{H}}^S_g$ by the coil 8.
The behavior of the neutron spin for the case of
$\alpha=90^\circ$ is shown in  Fig.~\ref{fig:4}. We use the same coordinate
system $(X,Y,Z)$ in Fig.~\ref{fig:4} and Fig.~\ref{fig:1}.

\begin{figure}[htbp]
	\centering
		\includegraphics[width=0.75\textwidth]{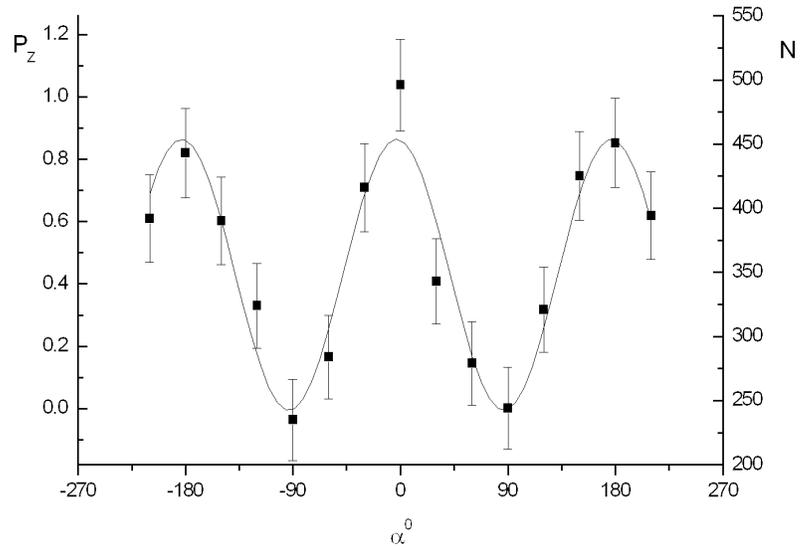}
\caption{The dependence of  the neutron intensity
$N$ on the angle $\alpha^\circ$ between the Schwinger magnetic field
${\bf{H}}^S_g$ and the vector of polarization of incident neutrons
for the Bragg angle $\theta_B=84^\circ$.}
\label{fig:5}       
\end{figure}

The dependence of a neutron beam intensity on the angle $\alpha$ 
after the analyzer 10 was studied to observe the depolarization effect
for diffracted neutrons.
The analyzer 10 transmits the neutrons with  the
polarization parallel to ${\bf{H}}^S_g$ only. The measurements are
similar to those, using a spin-echo
technique.

The neutron beam passed through the polycrystalline BeO filter 3 of a $120~$mm
thickness  to reduce the
contribution of the background reflections (see Fig.~\ref{fig:2})
to the forward diffraction beam.
The residual contribution of them  was estimated to be
$\simeq(20\pm10)\%$ of the useful intensity of neutrons
diffracted by the (110) plane. The uncertainty of this contribution
results in a systematic error of a measured value.

If the neutron spins  turn round  by the angles $\pm
\Delta\phi_0^S$ in the crystal for the waves described by $\psi^{(1)}$
and $\psi^{(2)}$, the counting rate $N$ of the neutrons
after the analyzer will be:
\begin{equation}
     N=N_0(1+P_Z)=N_0(1+P_0(\cos \Delta\phi_0^S \sin^2 \alpha+\cos^2
     \alpha)),
\label{eq:Na}
\end{equation}
where $P_Z$ is the projection of a neutron polarization after the crystal
on the direction ${\bf{H}}^S_g$. In the case of
$\Delta\phi_0^S=0$ we will have $P_Z\equiv P_0$ and
$N$ will not depend on the angle $\alpha$. The value of initial polarization
was $P_0=(87\pm3)\%$ for neutrons with the wavelength $\lambda \simeq
4.8$\AA.

The example of the dependence $N(\alpha)$ is shown in  Fig.~\ref{fig:5}.
The values of polarization $P_Z$ are shown on the left axis of ordinates.
The solid curve in Fig.~\ref{fig:5} is a result of
fitting the experimental points by the dependence (\ref{eq:Na}).

\begin{figure}[htb]
	\centering
		\includegraphics[width=0.75\textwidth]{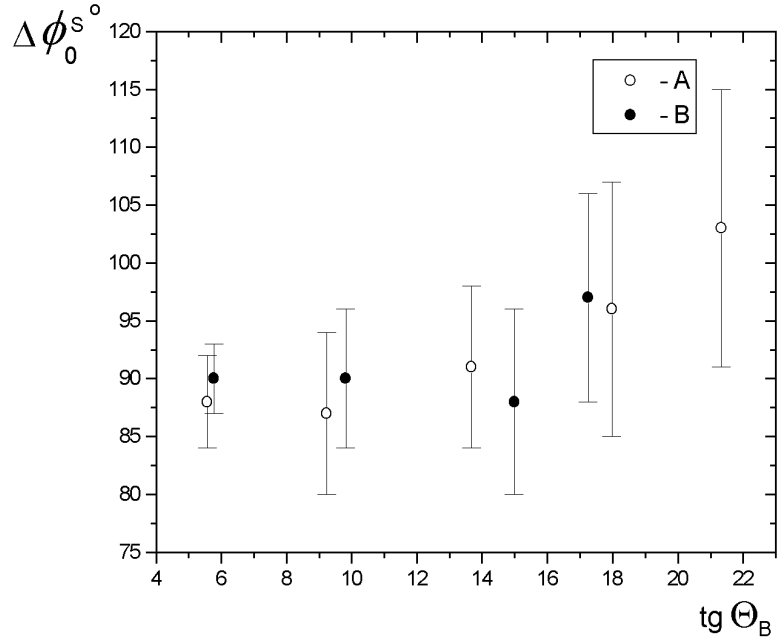}
\caption{The dependence of the angle $\Delta\phi_0^S$ of a neutron
spin rotation due to Schwinger interaction  on
the tangent of Bragg angle. {\bf A} and {\bf B} are
two crystal positions (see Fig.~\ref{fig:0}, and Fig.~\ref{fig:1}).}
\label{fig:6}       
\end{figure}

As it has been noted  earlier \cite{dedm,polart}, the angle of neutron
spin rotation due to Schwinger interaction does not depend on
the Bragg angle, and that is experimentally proved (see Fig.~\ref{fig:6}).

The experimentally observed result corresponds to the interplanar
electric field,
acting on a diffracted neutron, equal to
\begin{equation}
E_{(110)}=(2.24\pm 0.05(0.20))10^8~ \mbox{V/cm},
\end{equation}
The systematic error caused by uncertainty of the contribution of
background reflections is pointed in parentheses.

The experimental values are in good agreement with the earlier
theoretical predictions and confirm the opportunity to increase
more than by an order of magnitude the sensitivity of the method to
neutron EDM,
using the angles of diffraction close to $90^\circ$. It is
experimentally shown that the value $E\tau$ determining the
sensitivity of the method in our case can reach
$\sim 0.2\cdot 10^6$~V~s/cm, what is comparable with that of the UCN method
($\sim 0.6\cdot 10^6$~V~s/cm)\cite{edmlast} and much more than the value
obtained by
Shull and Nathans ($\sim 0.2\cdot 10^3$~V~s/cm)\cite{Shull}.

\section{Experimental test of the sensitivity}

In 2002 the test experiment was carried out at ILL (Grenoble, France)\cite{LDM_sens}.
The main purpose of this experiment  was to estimate
the statistical sensitivity of the Laue diffraction method for the neutron
EDM, using the available quartz crystal and the facility for particle physics with cold polarized neutrons
PF1A or PF1B at the ILL high flux reactor. The scheme of the experiment was similar to that of the previous one (see Fig.~\ref{fig:1}$^{a)}$). 
The TOF distance was $\approx 1$ m.
The experiment was carried out using the (110) plane
(d=2.456{\AA}) of a quartz crystal with the sizes 
$14\times 14\times 3.5$~cm$^{3}$ prepared and tested at PNPI (Gatchina, Russia). 
The mosaicity of the crystal was less than 1" over all crystal volume. 

\begin{figure}[htbp]
	\centering
		\includegraphics[width=0.85\textwidth]{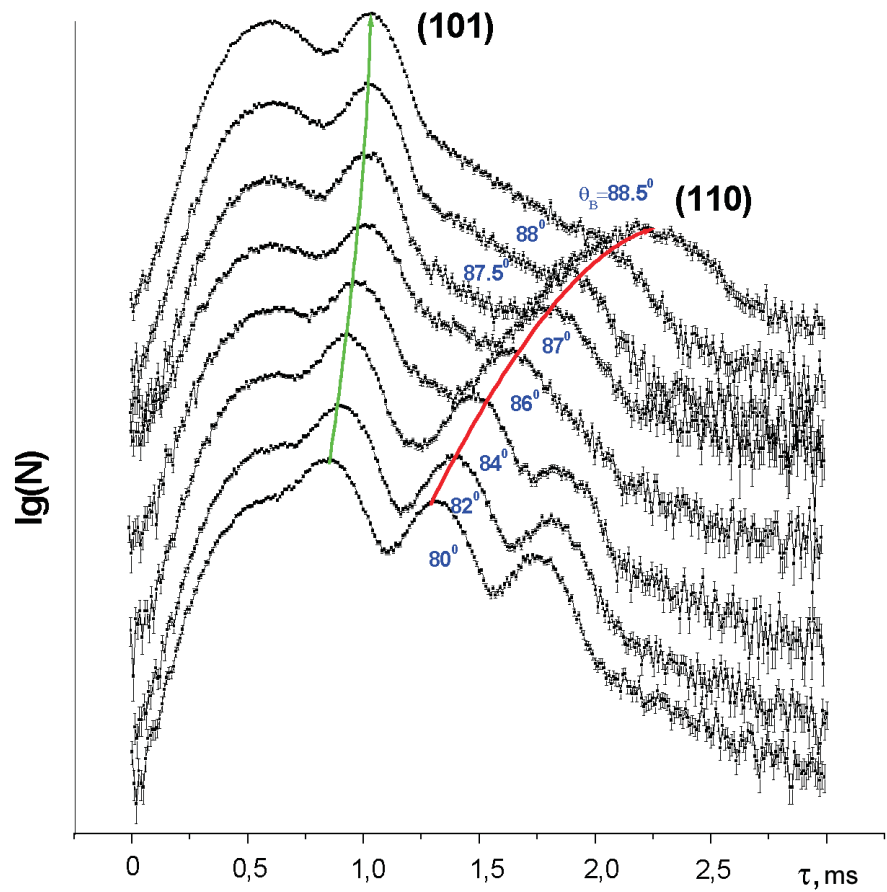}
    \caption{Spectrum of the forward diffracted neutron beam for different
      Bragg angles. Note the principal difference in the behavior of
      the working plane (110) and the other crystallographic planes (for
      instance, (101)). }
    \label{fig:TOF}
\end{figure}

Examples of experimental TOF spectra of the forward diffracted beam are
shown in Fig.\ref{fig:TOF}. The positions and widths of the different reflection peaks
coincide with the theoretical expectations.

\begin{figure}[htbp]
	\centering
		\includegraphics[width=0.85\textwidth]{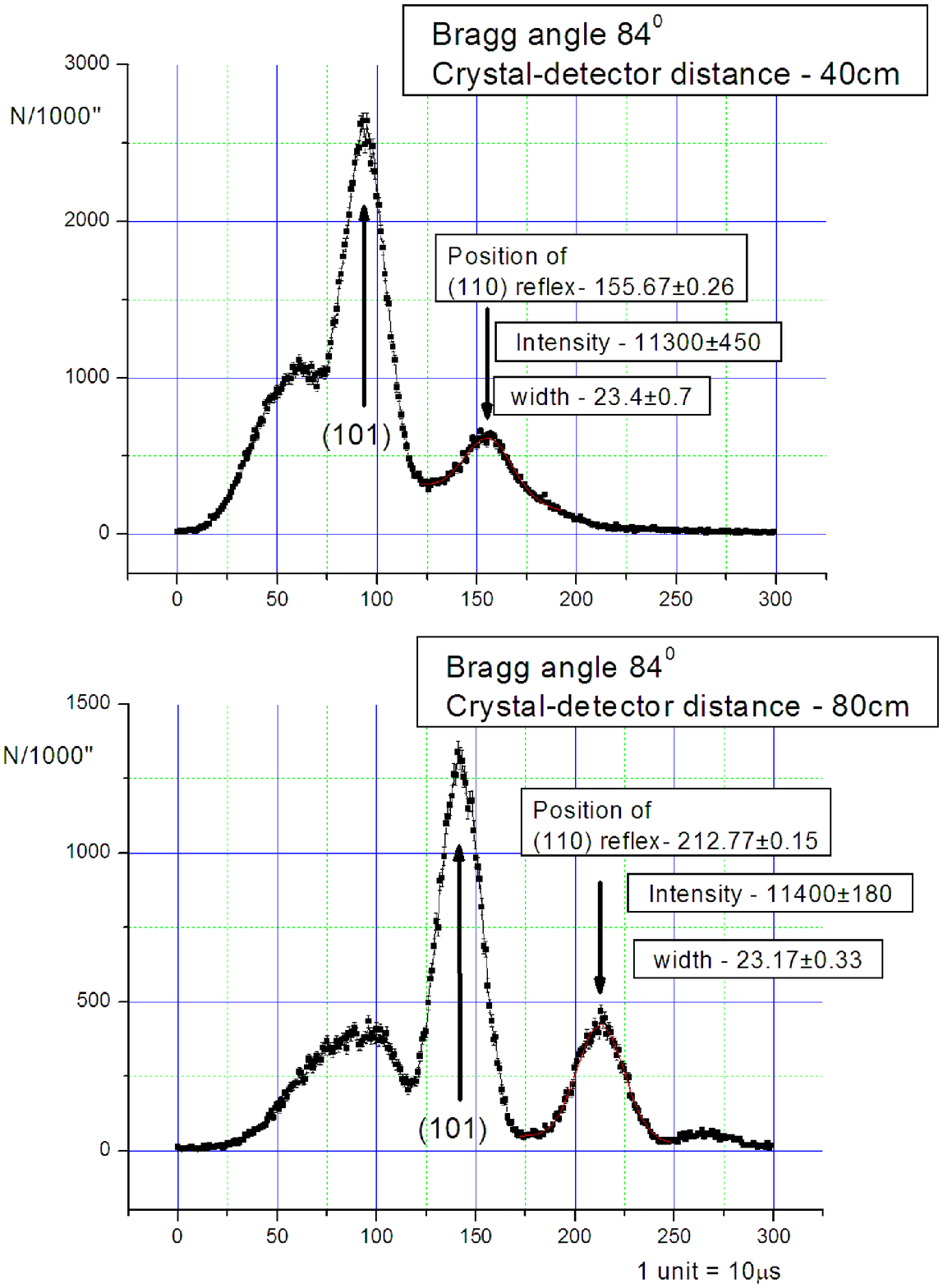}
    \caption{Spectrum of the forward diffracted neutron beam for two time of flight distances.}
    \label{fig:L_dist}
\end{figure}

One can see from Fig.\ref{fig:TOF} that the reflection peaks are located on a
background. Its spectrum is proportional to that of the incident beam.
This background is due to incoherent scattering
of the incident neutrons inside the crystal. It can be 
reduced essentially by increasing the distance between the crystal and detector
(see Fig.\ref{fig:L_dist}) and will
be negligible for the final geometry of the experiment.

\begin{figure}[htbp]
	\centering
		\includegraphics[width=0.70\textwidth]{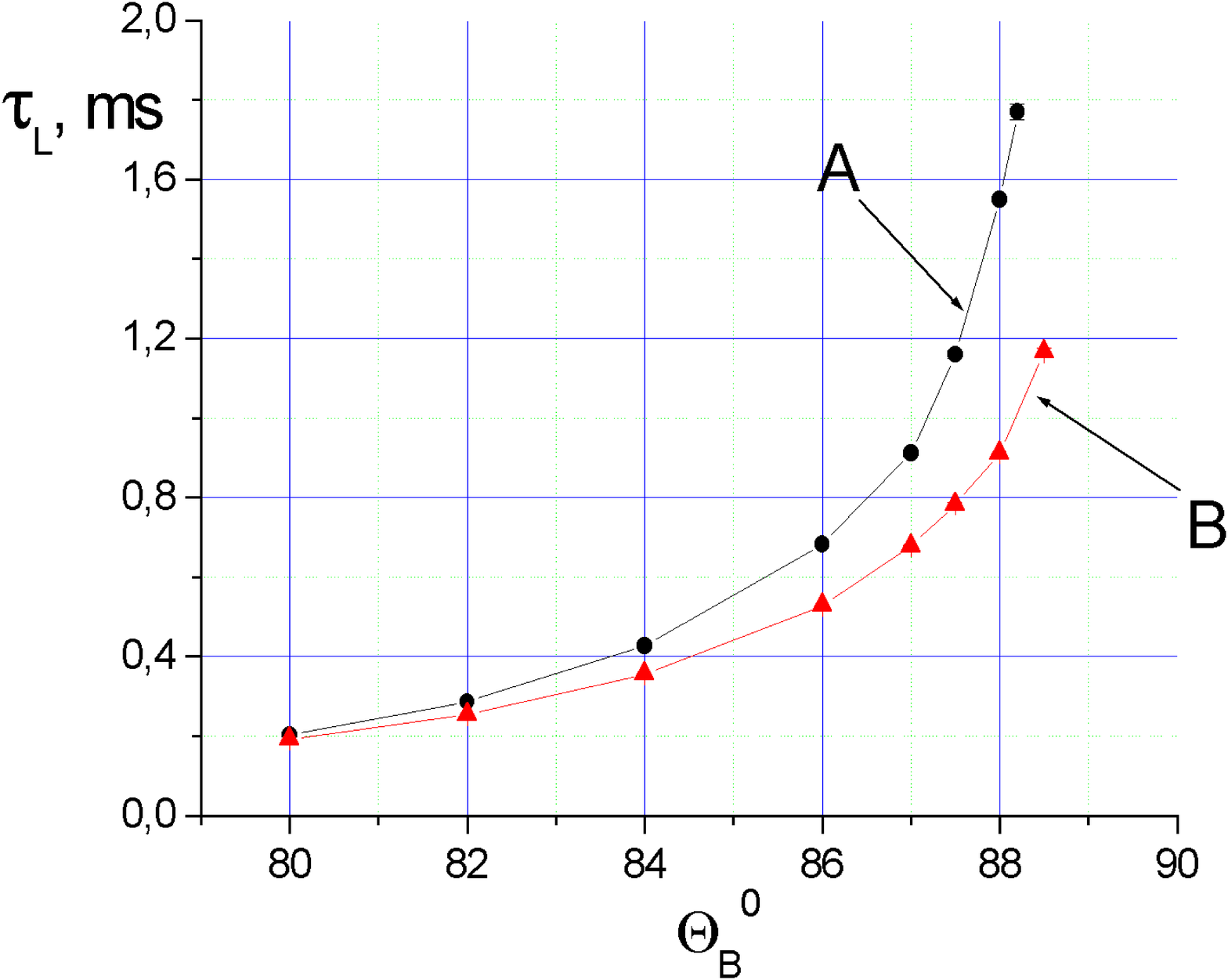}
    \caption{The dependence of the time of neutron stay in the crystal on 
      the Bragg angle.}
    \label{fig:Graph22}
\end{figure}

The measured dependencies of the time the neutrons stay inside the crystal are
shown in Fig.\ref{fig:Graph22}. The two curves correspond to the symmetric
crystal positions A and B with the same Bragg angles, see Fig.~\ref{fig:1}$^{a)}$.
The difference of these two curves is due to an inaccuracy of the initial
angular position of the crystal. The time of stay reaches 1.8~ms for the
Bragg angle
$\theta _{B} \approx 88.5^\circ$ which corresponds to the mean neutron
velocity $v_\parallel\approx 20$~m/s in the crystal. The incident neutron
velocity was $ \approx 800$~m/s ($\lambda  \approx  5$~\AA).

\begin{figure}[htbp]
    \begin{center}
        \includegraphics[width=0.70\textwidth]{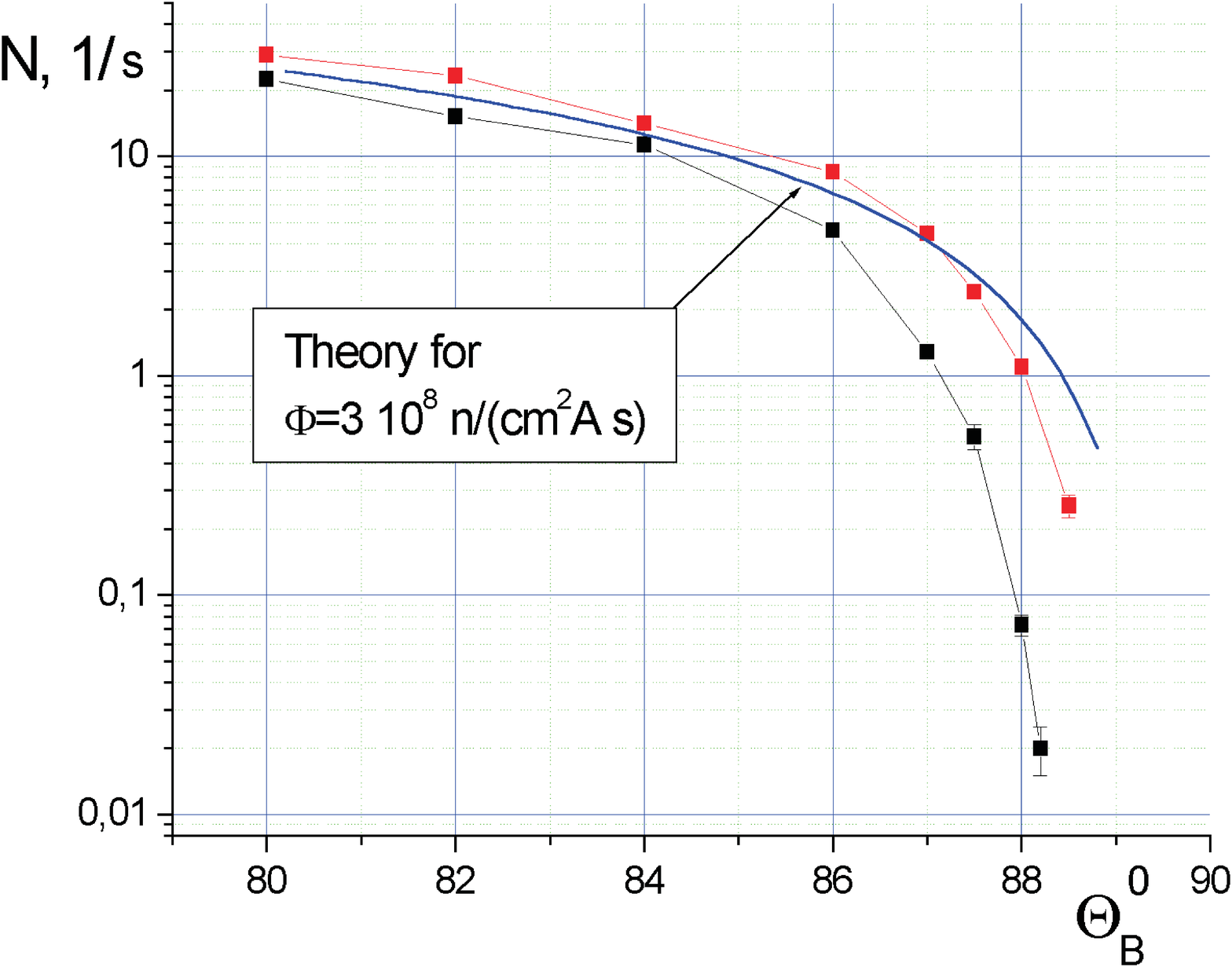}
    \end{center}
    \caption{Dependence of the intensity of the (110) reflection of the quartz
      crystal on the Bragg angle. The two curves correspond to the symmetrical
      crystal positions A and B (see  Fig.~\ref{fig:1}$^{a)}$).}
    \label{fig:Graph23}
\end{figure}

The dependence of the (110) reflection intensity  on the Bragg angle is shown
in Fig.\ref{fig:Graph23}. One can see a good coincidence of the theory with the
experimental data for Bragg angles smaller than 86$^\circ$. The disagreement
between theory and experiment for Bragg angles larger than 87$^\circ$ can be
explained by the decrease of the area of the incident neutron beam "`seen"' by 
the crystal for large angles of diffraction.

\begin{figure}[htbp]
    \begin{center}
        \includegraphics[width=0.70\textwidth]{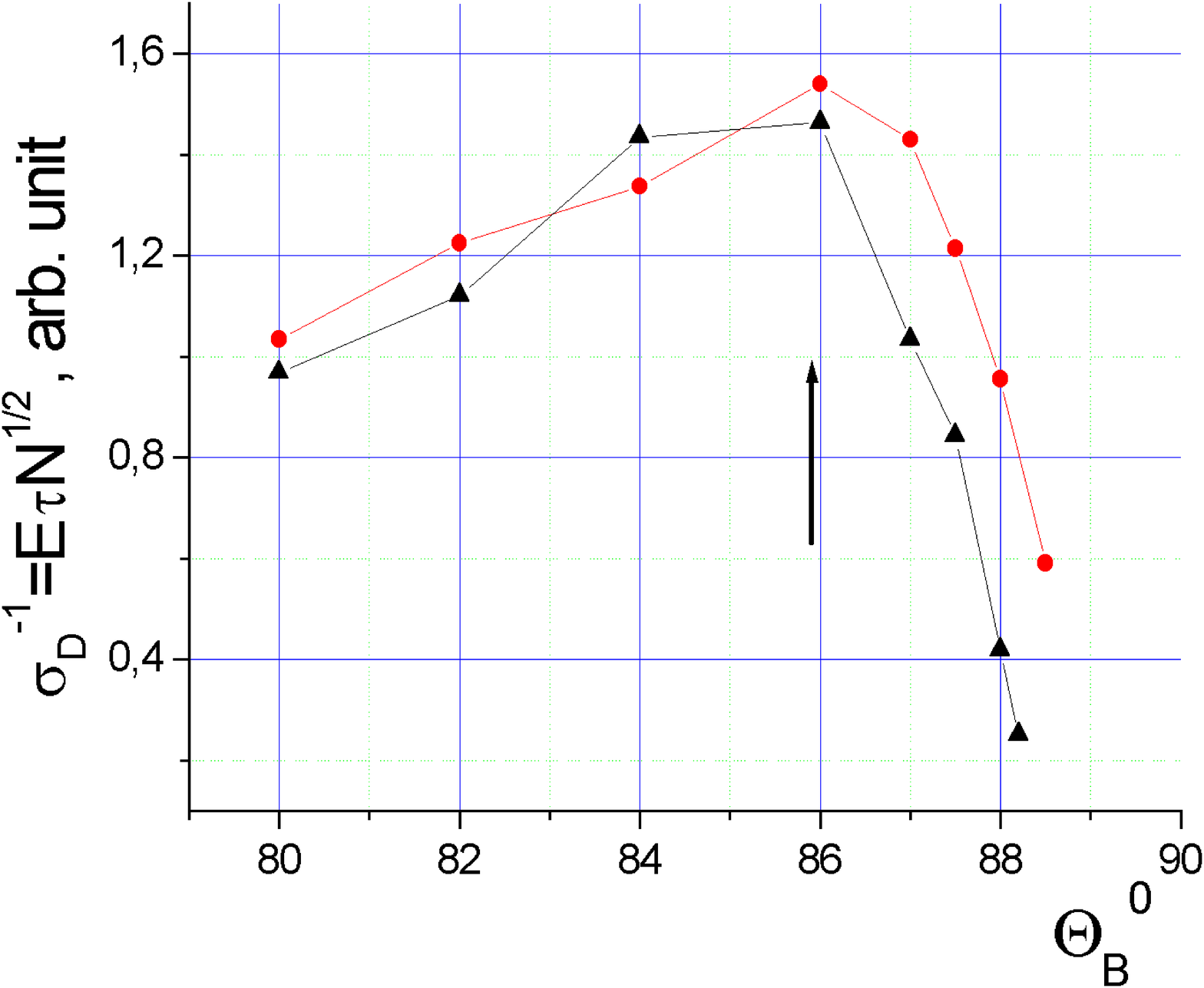}
    \end{center}
    \caption{The measured dependence of the sensitivity of the method for the
      neutron EDM on the Bragg angle.}
    \label{fig:Graph24}
\end{figure}

These measurements allow us to determine the dependence of the statistical
sensitivity 
for the neutron  EDM on the Bragg angle (see Fig.\ref{fig:Graph24}). This
dependence has maximum for the Bragg angle equal to 86$^\circ$.

Using the experimentally measured values of the  time of neutron passage through the
crystal (Fig.\ref{fig:Graph22}), the intensity of the diffracted beam
(Fig.\ref{fig:Graph23}), and
the previously measured value of the electric field for the (110) crystallographic plane 
($E_{g}=2.2\cdot 10^{8}$~V/cm \cite{dptfe,PhysB2001}) we can estimate the
statistical sensitivity of this method to measure the neutron EDM, using the world's
highest intensity cold neutron beam PF1B and the quartz crystal. This sensitivity is equal to
$\sim 6 \cdot 10^{-25}~e\cdot {\rm cm}$ per day for the Bragg angle
$\theta _{B}= 86^\circ$. 

The comparison with the
magnetic resonance method using ultracold neutrons (UCN-method)
\cite{pnpiedm,illedm,edmlast} is given in Table \ref{Tab:Sens}.

\begin{table}[htbp]
\caption{
The comparison of the Laue diffraction method with the UCN one.
The intensity for the Laue diffraction scheme is recalculated from the experimental value (see Fig. \ref{fig:Graph23}) for the quartz crystal
dimensions $3.5 \times 14 \times 25$ cm$^3$ and Bragg angle $\theta_B=86^\circ$
($\pi/2 -\theta_B\simeq 1/15$).}

\begin{center}
\begin{tabular}{|c|c|c|}
\hline \hline
        &  UCN-method \cite{edmlast}  & Laue diffraction  \\
        &                     &  method  \\ \hline \hline
  $E$ (kV/cm) &       4.5      &             2.2 $\cdot 10^5$\\ \hline
    $t$ (s) &   130  &  0.7 $\cdot 10^{-3}$  \\
     &   ($v=$5-6 m/s) &   ($v_{\parallel}\approx 50$ m/s) \\ \hline
$Et$ (kV s/cm)  &    585        &                150\\ \hline
$N$ (neutrons/s) &       60     &   $1\cdot 10^3$   \\ \hline\hline
$\sigma_D$ e$\cdot$cm per day  & $6\cdot 10^{-25}$ & $6\cdot 10^{-25}$ \\ \hline\hline
\end{tabular}
\end{center}
          \label{Tab:Sens}
\end{table}

Possible further progress of this method for the neutron EDM search experiment may be associated with the use of some other  crystals. In principle, there are the crystals that can allow to increase the method sensitivity about by an order of value in comparison with the quartz one (see Table.~\ref{Tab:Cryst}).

      \begin{table}[htbp]
      \caption{Parameters of some noncentrosymmetric crystal suitable for the EDM experiment.
      $\tau_a$ is a time of neutron life in the crystal (time of absorption).}
      \begin{center}
      \begin{tabular}{|c|l|l|l|c|c|c|}
      \hline\hline
      Crystal&Group &$hkl$&$d,$ (\AA)&$E_g$,&$\tau_a$,
      &$E_g\tau_a$,\\
           &Symmetry &  & & 10$^9$V/cm &  ms  &(kV s/cm)\\
      \hline\hline
      $\alpha$-quartz & 32($D^6_3$)&111&2.236& 0.23 & 1.0 &230\\
       (SiO$_2$)   &            &110& 2.457& 0.20 &   & 220\\
      \hline
       Bi$_{12}$SiO$_{20}$&$123$&433& 1.75& 0.43 & 4.0 & 1720 \\
          &           &312&2.72& 0.22 &    &880 \\
      \hline
        Bi$_{4}$Si$_{3}$O$_{12}$  &$-43m$ & 242&2.1 &0.46 & 2&920\\
             &            & 132&2.75 &0.32&    &640 \\
      \hline
       PbO &$Pca21$ & 002&2.94& 1.04&1&1040\\
             &            & 004&1.47 &1.0 &   &1000 \\
      \hline
       BeO  &$6mm$              & 011&2.06 & 0.54 & 7.0 &3700 \\
             &            & 201&1.13 & 0.65 &   &4500 \\
      \hline\hline
      \end{tabular}
      \end{center}
          \label{Tab:Cryst}
 \end{table}

\newpage
\section{Neutron optics in  noncentrosymmetric crystal}

Here we consider the neutron-optic effects for a neutron, moving
through a noncentrosymmetric crystal with the energy and direction
far from the Bragg ones, when the deviation from the exact Bragg
condition reaches $(10^3 - 10^5)$ Bragg widths.

\begin{figure}[htb]
	\centering
		\includegraphics[width=0.65\textwidth]{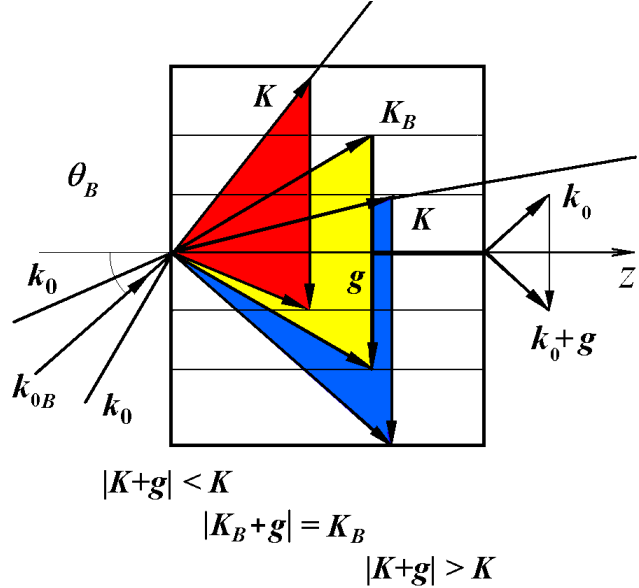}
\caption{Neutron movement with the different directions 
of the wave
vectors $\mathbf{K}$ with the respect to reciprocal lattice
vector $\mathbf{g}$. a) $|\mathbf{K+g}|> K$,
neutrons are concentrated at the "nuclear planes"
(maxima of nuclear potential). b) $|\mathbf{K+g}|< K$,
neutrons are concentrated between the "nuclear planes". These two
cases correspond to neutron optics. c) The third case
$|\mathbf{K+g}|= K$ corresponds to the neutron diffraction,
when the both kinds of waves excite in the crystal.}
\label{difr:1}         
\end{figure}

\begin{figure}[htbp]
	\centering
		\includegraphics[width=0.65\textwidth]{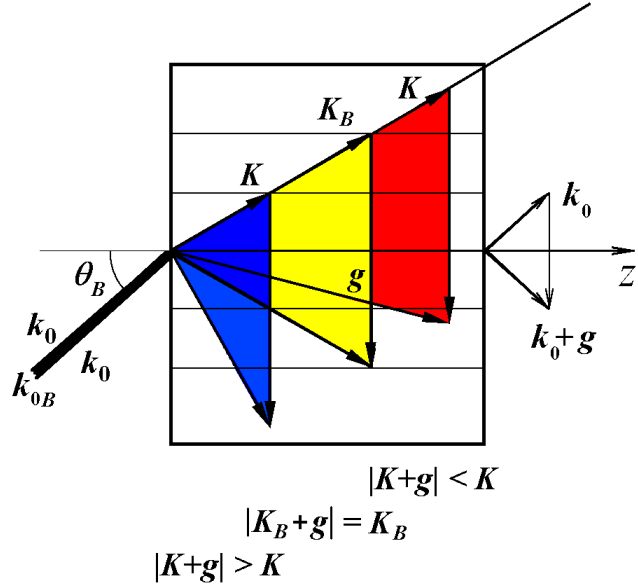}
\caption{
Neutron movement with the different
absolute values of the wave
vectors $\mathbf{K}$ (different wave lengths).
The same three cases
as in the Fig.~\ref{difr:1} $|\mathbf{K+g}|> K$, $|\mathbf{K+g}|< K$ and
$|\mathbf{K+g}|= K$ }
\label{difr:2}       
\end{figure}

The essence of the phenomenon is the following. Let a
neutron is moving through the crystal and the Bragg condition is not
satisfied for any crystallographic plane.
In this case the distribution of the neutron density
$|\psi({\mathbf r})|^2$ in the crystal can be written using a
perturbation theory \cite{dedm2}:
\begin{equation}
|\psi({\mathbf r})|^2=1+\sum_g \frac{2 v_g}{E_k-E_{k_g}}\cos ({\bf g r}+\phi_g).
\end{equation}
where $E_k=\hbar^2 k^2/2m$ and $E_{k_g}=\hbar^2 |{\bf k}+{\bf g}|^2/2m$
are the energies of the neutron with the wave vectors  ${\mathbf k}$
and ${\mathbf k}+{\mathbf g}$ in the crystal, the values $v_g$ and $\phi_g$ are respectively the absolute magnitude 
and the phase of the $g$-harmonics amplitude $V_g=v_g\exp{i\phi_g}$ of the neutron crystal interaction   potential, which  has a form 
\begin{equation}
V({\mathbf r})=
\sum_g V_g\exp{i(\bf g r)}=V_0+\sum_g 2v_g\cos{({\bf g r}+\phi_g)},
\end{equation}
difference $\Delta_g=E_k-E_{k_g}$  describes the deviation from the Bragg condition
measured in the energy units.
One can see that the neutrons are concentrating either on
the maxima or on the minima of
the periodic potential,
depending on the sign of  $\Delta_g$ (see Fig. \ref{difr:1},\ref{difr:2}), the degree of this concentration
being determined by the value of $v_g/\Delta_g$.

The neutron interacts with the nonmagnetic crystal by
the nuclear forces, therefore the neutrons will concentrate
on the maxima (or on  the minima) of the {\it nuclear} crystal potential,
so the neutrons in the noncentrosymmetric crystal will  move
under electric interplanar field as in the diffraction case  analyzed above.
The sign of the electric field  depends on the sign of the
deviation parameter $\Delta_g=E_k-E_{k_g}$.\footnote{The equality $E_k=E_{k_g}$
corresponds to exact Bragg condition for the plane system
${\bf g}$. In this case the perturbation theory (valid for $v_g\ll (E_k-E_{k_g})$)
becomes inapplicable, and
one should use the two wave diffraction theory.}

Due to such concentration the neutron kinetic
energy in the crystal $E_k=\hbar^2 k^2/{2m}$ (in the second order of the perturbation theory) will be equal to
\begin{equation}
{\hbar^2 k^2\over{2m}} ={\hbar^2 k_0^2\over{2m}}-V_0-\sum_g \frac{V_g V_{-g}}{\Delta_g}\equiv{\hbar^2 K^2\over{2m}}-\sum_g \frac{V_g V_{-g}}{\Delta_g},
 \label{Ek}
\end{equation}
where $k_0$ is the wave vector of an incident neutron and $k$ is the wave vector of the neutron in the crystal, $K$ is the wave vector in the crystal with the mean nuclear potential $V_0$ taken into account $\hbar^2 K^2/2m=\hbar^2 k_0^2/2m-V_0$.
As follows from (\ref{Ek}),  the neutron, moving through the crystal
far from the Bragg directions, nevertheless, "feels" the crystal structure.

For the case of non-magnetic and non-absorbing crystal
the expression for $V_g$ can be written as \cite{dedm2}:
\begin{equation}
  V_g=v^N_g e^{i\phi^N_g} + i v^E_g e^{i\phi^E_g}
\mu \frac{\mbox{\boldmath $\sigma$}[{\mathbf g}\times{\mathbf v}]}{c},
\end{equation}
where $v^N_g, \phi^N_g$ are respectively the absolute magnitude  and the phase of the
$g$-harmonics amplitude of the nuclear neutron-crystal potential,
$v^E_g, \phi^E_g$ are the absolute magnitude  and the phase of the $g$-harmonics amplitude
of the electric potential of crystal,
$\mu, v$ are the magnetic moment and  the velocity of  neutron, $c$
is the light speed.

By substituting this expression into (\ref {Ek}) and  taking into account that
for non-absorbing crystal $V_g=V_{-g}^*$, we shall
obtain
 
\begin{equation} 
  {\hbar^2 k^2\over {2m}} ={\hbar^2 K^2\over{2m}}-\sum_g \frac{(v^N_g)^2}{\Delta_g}-
 \mu \frac{\mbox{\boldmath $\sigma$}[{\mathbf E}_{sum}\times{\mathbf v}]}{c},
\end{equation}
where
\begin{equation}
{\mathbf E}_{sum}=\sum_g \frac{2 v^N_g}{\Delta_g}
 v^E_g{\mathbf g} \sin(\Delta \phi_g)
\label{Esum}
\end{equation}
is a resultant electric field affecting a neutron in the crystal.
Here $\Delta \phi_g\equiv \phi^N_g-\phi^E_g$ is
the phase shift between  $g$-harmonics  of nuclear and electric potentials of the crystal.

We should note
that the neutron refraction index $n$  in this case will depend on the neutron spin direction
\begin{equation} 
 n^2= k^2/k^2_0=n_0^2 -\Delta n_d^2-\Delta n_s^2 ,
 \end{equation}
  where $n_0^2=K^2/k_0^2$ is a  square of a mean   refraction index for neutron in the crystal, 
\begin{equation} 
  \Delta n_d^2 = (2m/\hbar^2k_0^2) \sum_g (v^N_g)^2 /\Delta_g
  \end{equation}
   is a small diffraction correction to the square of mean refraction index, 
   \begin{equation} 
\Delta n_s^2 = (2m/\hbar^2k_0^2)(\mu \mbox{\boldmath $\sigma$}[{\mathbf E}_{sum}\times{\mathbf v}])/c
  \end{equation}
 is a spin dependent correction arisen due to interference of nuclear and electric amplitudes.

For the centrosymmetric crystal $\Delta \phi_g\equiv 0$ and
therefore  ${\mathbf E}_{sum}\equiv 0$.
In the noncentrosymmetric crystals there are the crystallographic planes, 
for which  $\Delta \phi_g\neq 0$ and so the
electric field acting on a neutron will be nonzero
${\mathbf E}_{sum} \neq 0$. 

Therefore a spin dependence arises for the neutron-crystal
interaction, which  leads to  different values of a neutron wave vector in the crystal for two opposite spin orientations that in turn leads to neutron spin rotation
around  the direction of Schwinger magnetic field
${\mathbf H}^S_{sum}=[{\mathbf E}_{sum}\times{\mathbf v}]/{c}$. 

The
rotation angle for the crystal length $L$ will be equal
\begin{equation}
\Delta \varphi_s= \frac{2\mu}{\hbar }
 \frac{\mbox{\boldmath $\sigma$}[{\mathbf E}_{sum}\times{\mathbf v}]}{c} \frac{L}{v}.
\label{dfi}
\end{equation}

One should add an imaginary part into the nuclear crystal potential
to describe the absorbing crystal:
\begin{equation}
V_g=v^N_g e^{i\phi^N_g} + i v^{N'}_g e^{i\phi^{N'}_g} + i v^E_g e^{i\phi^E_g}
\mu \frac{\mbox{\boldmath $\sigma$}[{\mathbf g}\times{\mathbf
v}]}{c}.
\end{equation}
Here $v^{N'}_g, \phi^{N'}_g$ are the amplitude and phase of
$g$-harmonics
of the imaginary part of the nuclear potential.

The value of kinetic energy in the crystal becomes equal to:
\begin{equation}
{\hbar^2 k^2\over {2m}} ={\hbar^2 K^2\over{2m}}-V_{\langle g\rangle}-i( V'_0+V'_{\langle g\rangle})-
\mu \frac{\mbox{\boldmath $\sigma$}[({\mathbf E}_{sum}+i {\mathbf E'}_{sum})\times{\mathbf v}]}{c},
\label{Etild}
\end{equation}
where
\begin{equation}
V_{\langle g\rangle}=\sum_g \frac{(v^N_g)^2-(v^{N'}_g)^2}{\Delta_g},
\label{V0g}
\end{equation}
\begin{equation}
V'_{\langle g\rangle}=\sum_g \frac{2 v^N_g
v^{N'}_g\cos(\phi^{N}_g-\phi^{N'}_g)}{\Delta_g},
\label{v'}
\end{equation}
\begin{equation}
{\mathbf E'}_{sum}=\sum_g \frac{2 v^{N'}_g}{\Delta_g}
 v^E_g \sin(\phi^{N'}_g-\phi^{E}_g){\mathbf g}.
\label{E'}
\end{equation}

The estimations give that the
values of the diffraction corrections to a mean potential
for  $\alpha$-quartz crystal are $V_{\langle g\rangle}+iV'_{\langle g\rangle}\approx 10^{-3}(V_0+i V'_0)$,\\
$\mu \mbox{\boldmath $\sigma$}[({\mathbf E}_{sum}+i {\mathbf E'}_{sum})\times{\mathbf v}]/c\approx
10^{-6}(V_0+i V'_0)$ for the wide range of the incident neutron
wavelengths and sharply increase near the Bragg conditions. We should note also that in spite of a smallness the last correction  leads to relatively large and observable effects due to its spin dependence.

\section{Observation of the neutron spin rotation}
\label{sec:2}
\begin{figure}[htb]
	\centering
		\includegraphics[width=0.85\textwidth]{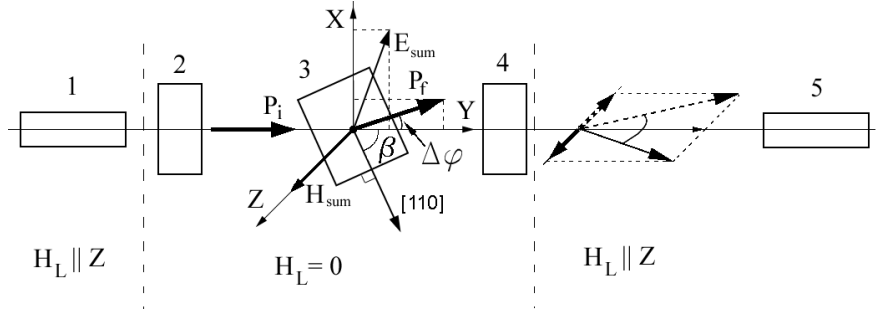}
\caption{Scheme of the experiment. 1 is a polarizer; 2 is a $\pi
/2$ coil to turn spin around X axis; 3 is the $\alpha$-quartz single crystal with
the sizes $14\times 14\times 3.5$~cm$^3$; 4 is a $\pm \pi /2$ coil
to turn around Y axis; 5 is an analyzer. ${\mathbf H_L}$ is the guiding
magnetic field; $O$ is an axis (parallel to $Z$ direction) of a crystal rotation;
${\mathbf P_i}$ and ${\mathbf P_f}$ are the polarizations of
neutron beam before the crystal and after it.}
\label{fig:7}       
\end{figure}

The experiment was carried out at the PNPI WWR-M reactor. 
Scheme of the experiment is shown in Fig.~\ref{fig:7}, see \cite{Appl_Ph_Srot,srjetpl,Nopt_NIMB}.

Initially the neutron spin was directed
along the neutron velocity (axis Y).  The X-component of the
polarization was measured after neutron passage through
the crystal. This component should be
equal to zero, if the spin rotation effect is absent.
Time of flight technique was used for measuring the
spectral dependence of polarization. The crystal was
 overturned around Z axis to eliminate the false effect
due to nonzero value of the X-component of polarization for real
setup.
 The effect changes its sign due to a change
 of the sign of the electric field in this case.

The measurement was carried out using the $\alpha$-quartz crystal
with the dimensions $14\times 14\times 3.5$~cm$^3$.
The crystal orientation was determined by the angle $\beta$
between the neutron velocity (Y axis) and the [110] crystal axis.

\begin{figure}[htb]
	\centering
		\includegraphics[width=0.75\textwidth]{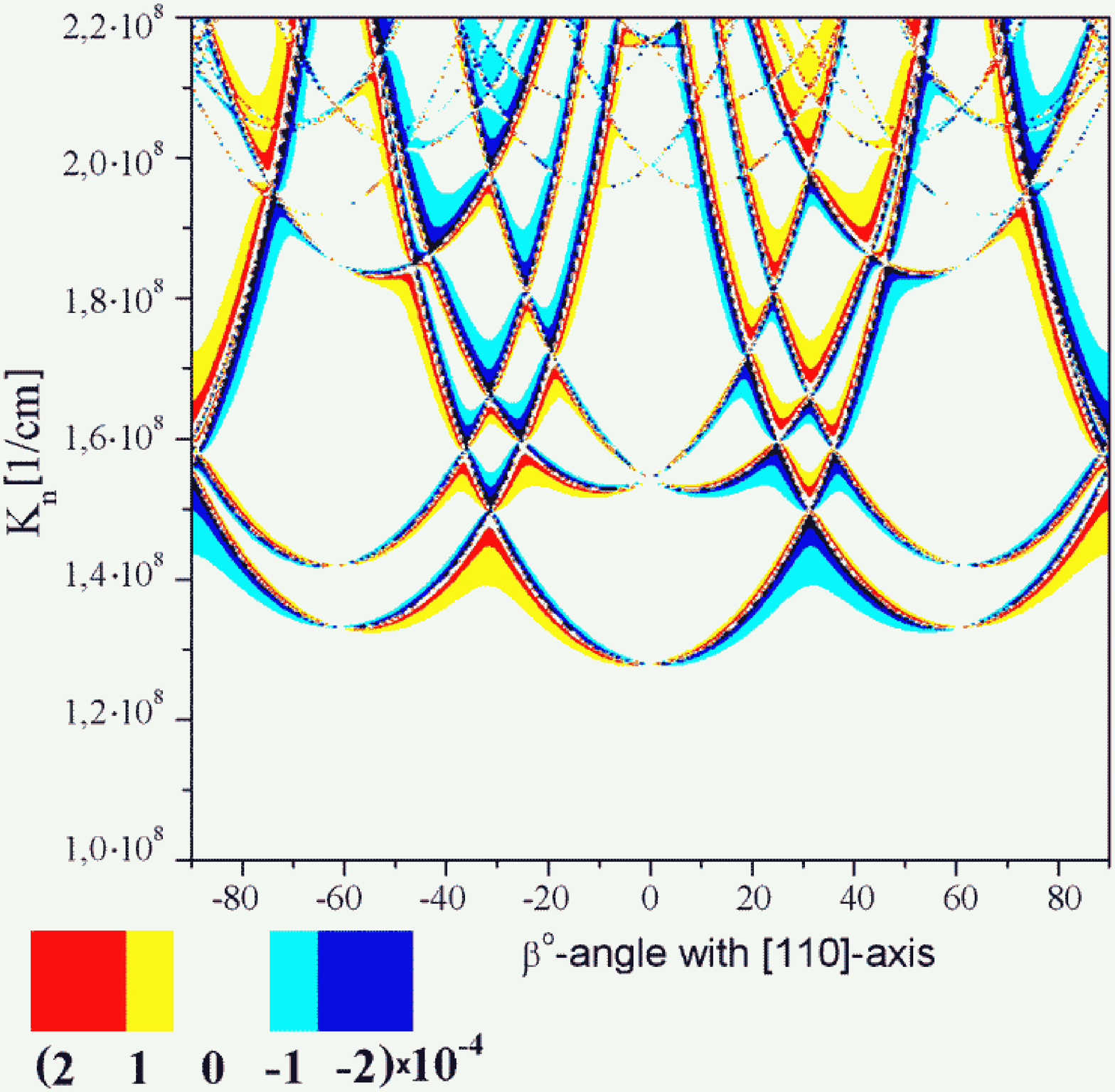}
\caption{Theoretical dependence of $\Delta \varphi_s$ on the wavevector and
direction of neutron for the $\alpha$-quartz crystal.}
\label{fig:dfis}       
\end{figure}

The theoretical dependence of the angle of neutron spin rotation
$\Delta \varphi_s$ on the value and direction of neutron
wavevector is shown in Fig.~\ref{fig:dfis}.

\begin{figure}[htb]
	\centering
		\includegraphics[width=0.75\textwidth]{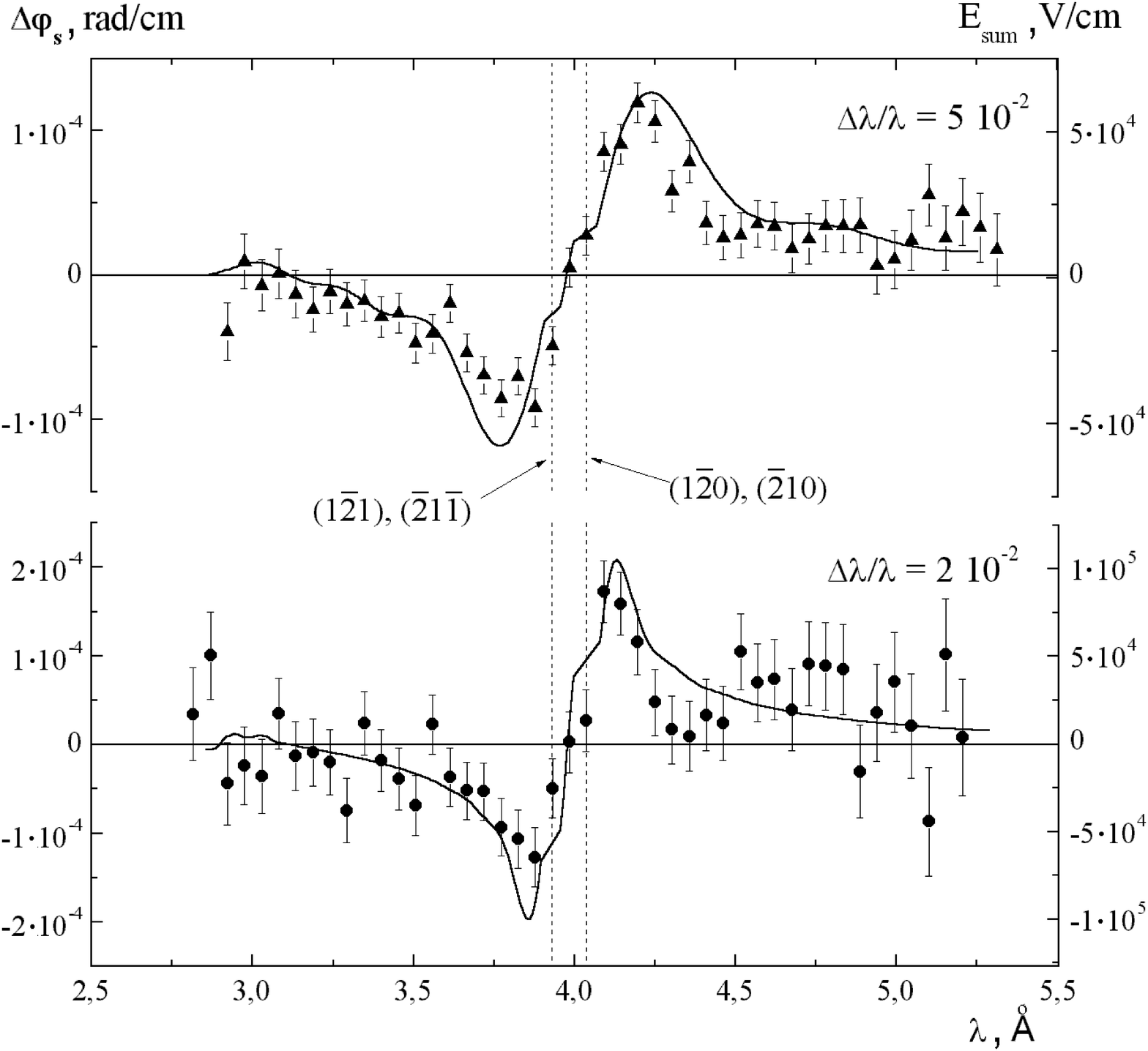}
\caption{Energy dependence of $\Delta \varphi_s$ for $\beta=90^\circ$. Solid curves are the theoretical dependence (\ref{dfi}) after averaging over the energy resolution of the experiment.}
\label{fig:9}       
\end{figure}

\begin{figure}[htb]
	\centering
		\includegraphics[width=0.75\textwidth]{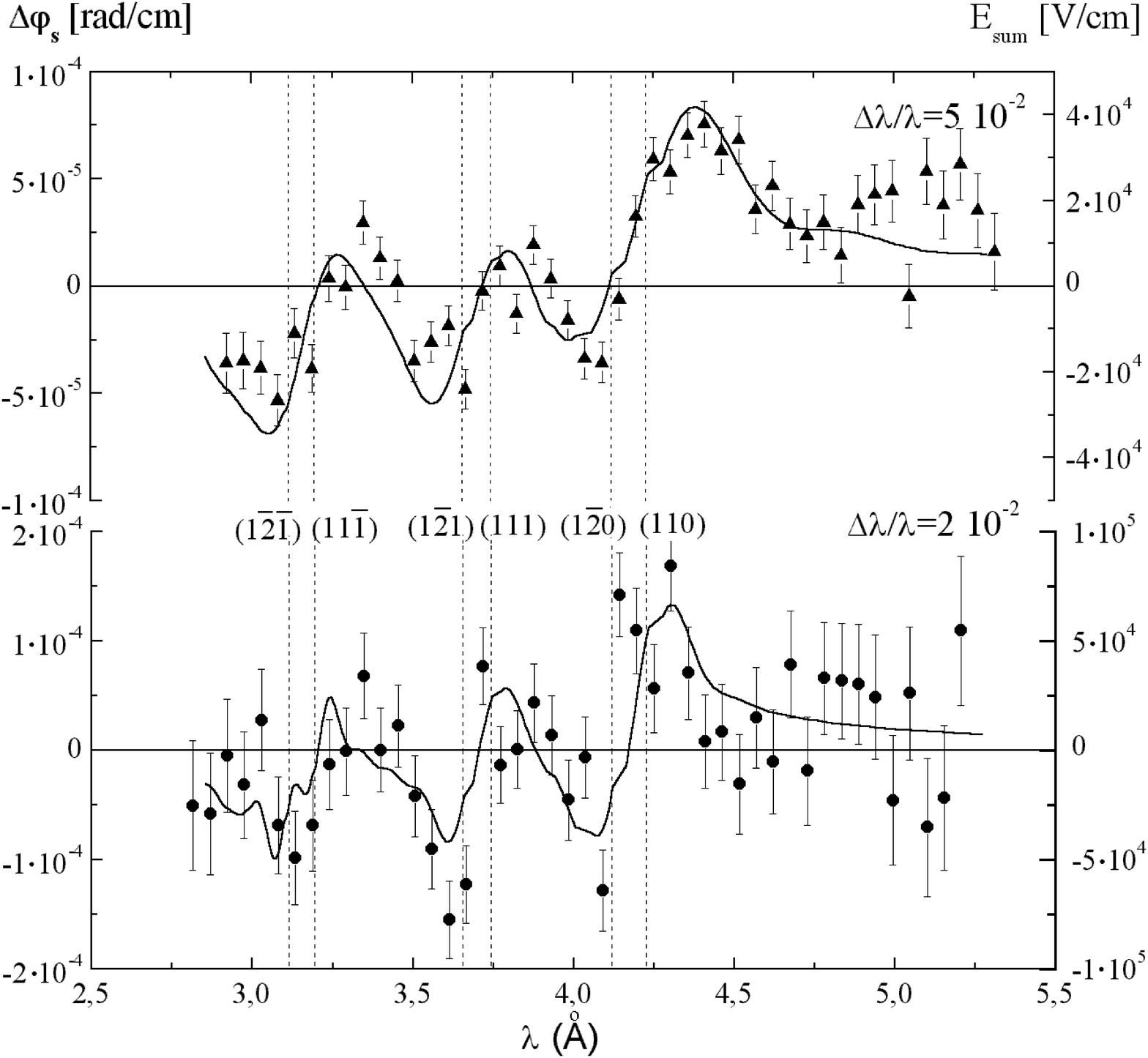}
\caption{Energy dependence of $\Delta \varphi_s$ for $\beta=30^\circ$.}
\label{fig:10}       
\end{figure}

The experiment was carried out for two crystal positions with
$\beta=90^\circ$ and $30^\circ$. The results are shown in
Figs.~\ref{fig:9},~\ref{fig:10}. Two plots at the figures correspond
to different energy resolution of the experiment
($\Delta\lambda/\lambda=
5\cdot10^{-2}$ and $=2\cdot10^{-2}$). The solid curves reproduce the
theoretical dependence (\ref{dfi})  averaged over energy
resolution. The dotted lines indicate the positions of
the crystallographic planes with nonzero value of $\Delta \phi_g$
(see (\ref{Esum})). One can see a good agreement between
theoretical and experimental results.  The resultant electric
field $|{\mathbf E}_{sum}|$ is shown on the right ordinates axis.
Its value is  $|{\mathbf E}_{sum}|\approx (1-10)\cdot
10^4$~V/cm for any point of spectrum, and so $\Delta
\varphi_s$ may reach $\pm 2\cdot 10^{-4}$~rad/cm.

\section{Spin rotation effect   for  Bragg reflected neutrons from the deformed crystal part}

In the previous chapter  we have considered the effect of neutron spin rotation for large deviations from exact Bragg condition ($ \sim 10^3 - 10^4$ Bragg width). The measured
effect has coincided with the theoretical one but has a small value \cite{srjetpl}. Here we  consider the situation when we can tune the sign and value of the deviation from the exact Bragg condition within a few Bragg width. In this case the  neutron spin rotation angle essentially increases due to increasing the electric field affected the neutron.

Let's consider the\ symmetric Bragg diffraction scheme with the Bragg angles close to the right one. Neutrons fall
on the crystal in the given direction with the energy close to the Bragg one  for the
crystallographic plane~$g$. Deviation from the exact Bragg
condition is described by the parameter $\Delta_g=E_k-E_{k_g}$, where $E_k = \hbar^2k^2/2m$
and $E_{k_g}=\hbar^2|{\rm {\bf k}} + {\rm {\bf g}}|^2/2m$ are the energies of a neutron in the states $\vert
k\rangle $ and $\vert k+g\rangle $ respectively.

In this case the neutron wave function inside the crystal 
in the first order of perturbation theory can be written \cite{dedm2}
\begin{equation}\label{Psi}
\psi ({\rm {\bf r}}) = e^{ - i\;{\rm {\bf kr}}} + a \cdot e^{ -
i({\rm {\bf k}} + {\rm {\bf g}}){\rm {\bf r}}},
\end{equation}
where
\begin{equation}\label{a}
a = \frac{\left| {V_g } \right|}{E_k - E_{k_g }} = \frac{\left| {V_g } \right|}{\Delta_g}.
\end{equation}
Here $V_g $ is $g$-harmonic of interaction potential of neutron
with crystal. For simplicity we consider
the case $a \ll 1$, so we can use the perturbation theory.

The electric field affected the diffracted neutron will be
equal to \cite{dedm2}
\begin{equation}\label{E}
{\rm {\bf E}} = 2{\rm {\bf E}}_g \cdot a,
\end{equation}
where ${\bf E}_g$ is the interplanar electric field for the exact
Bragg condition.

One can see that the sign and value of the electric field (\ref{E})
are determined by the sign and value of deviation $\Delta_g$ from the exact Bragg condition,
therefore to have the given electric field and so the effect of neutron spin rotation
we should select from the whole beam the neutrons with the 
corresponding deviation parameter $\Delta_g$.

 The presence of the electric field
will lead to an appearance of the Schwinger magnetic field
\begin{equation}\label{Hgs}
  {\bf H}_S= 1/c[{\bf E}\times \mbox{\bf v}_{\|}].
\end{equation}
 The neutron spin will rotate around the
${\bf H}_S$ by the angle
\begin{equation}
\varphi _s = \frac{4\mu H_S L_c} {\hbar
\mbox{v}_ \bot},
\end{equation}
$L_c$ is the crystal thickness, $\mbox{v}_{\|}$ and $\mbox{v}_ \bot$ are the components of neutron velocity parallel and perpendicular to the crystallographic plane correspondingly.

The main idea of the experiment is the following. We use a small controlled variation of the  interplanar distance $\Delta d$ (caused by heating, for example) near the exit crystal edge.   So some part of neutrons passed through the crystal will reflect from this small crystal part. These back diffracted neutrons  have the deviation parameter for the main part of crystal determined by $\Delta d$ and so they  propagate under corresponding electric field in both directions there and back.
Thermal deformation of the crystal edge is used to create such variation of the  interplanar distance.

\begin{figure}[ht]
\centerline{\psfig{file=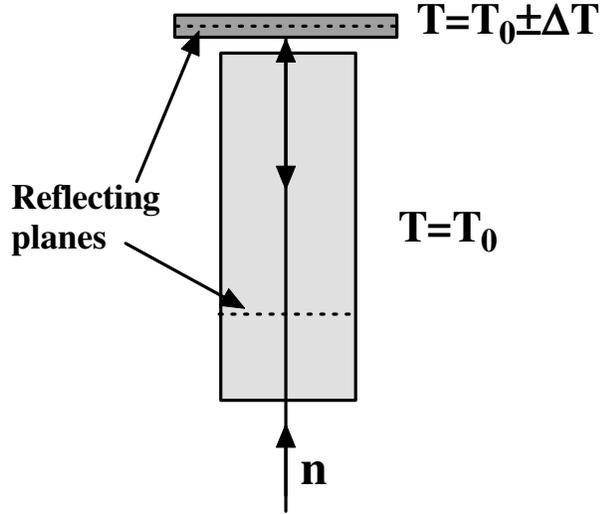,clip=,width=8 cm}}
\caption {Two crystals are in parallel position. Neutron reflected by the small crystal pass twice through the large crystal. The deviation parameter $\Delta_g $ for the large crystal is determined by the temperature difference $\Delta T$. }
\label{fig:1_1}
\end{figure}

We can use also two separate crystals in parallel position for this  purpose (see Fig.\ref{fig:1_1}). One can heat (or cool) the second small crystal. The part of neutrons passed through the first crystal with the corresponding  Bragg wave length will  reflect  by the second crystal with the given deviation parameter for the first (large) crystal. This deviation parameter will directly depend on the temperature difference between crystals.

Value of the wave length
Bragg width for (110) quartz plane ($d=2.45$\AA) is $\Delta \lambda _B / \lambda \approx
10^{ - 5}$. To shift the   wave length of the reflected neutrons by the one Bragg width we should heat (or cool) the second crystal to have the same value of $\Delta d / d$.
Linear coefficient
of the thermal expansion for quartz crystal is $\Delta L / L \approx 10^{ - 5}$ per degree.
Therefore, the deviation
$\pm \Delta \lambda _B $ corresponds to difference of the crystal
temperatures $\Delta T \approx \pm 1^0$.
We note that the different signs of this temperature difference will correspond to 
different signs of the electric field acting on the neutron.

\begin{figure}[ht]
\centerline{\psfig{file=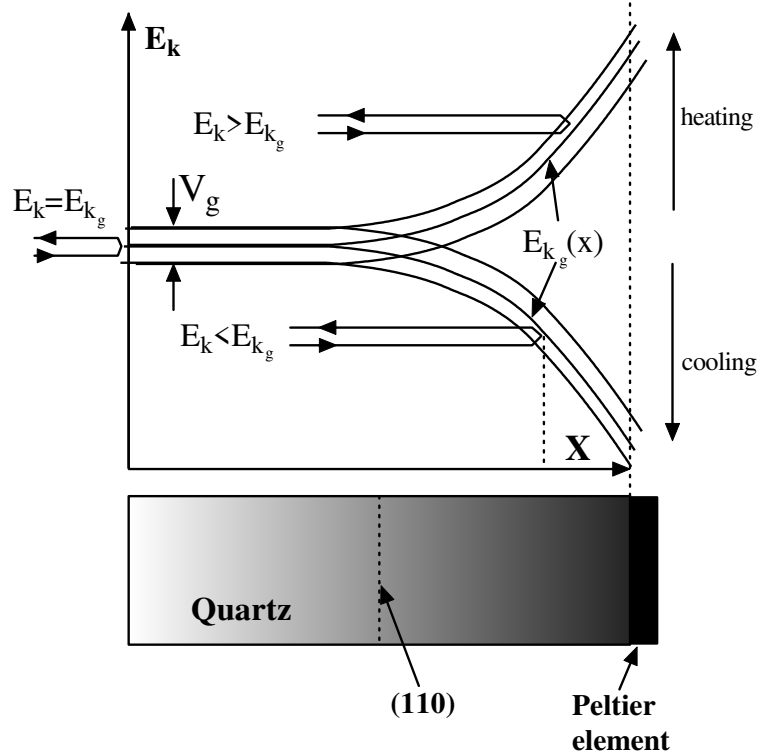,clip=,width=10 cm}}
\caption {Passage of the neutron through the crystal. Presence of the interplanar distance gradient result in forming the reflex near the back face of crystal.}
\label{fig:2_2}
\end{figure} 

Scheme of the neutron behavior in the crystal is shown in Fig.\ref{fig:2_2}.
Two samples of quartz crystal has been used in this experiment with the thicknesses along X axis $L_c=14$ and $27$ cm. The Peltier element has been attached to the back face of the crystal. That allows to create the temperature gradient in the crystal along the neutron trajectory. So the Bragg condition will vary along the neutron trajectory and different parts of crystal reflect the neutrons with the different $\lambda$. Therefore, the reflected beam  will contain not only the reflex from the entrance crystal face (corresponding to Bragg condition for $d$) but also the reflection from the back exit face (corresponding $d\pm \Delta d$) that twice pass through the crystal there and back. Moreover, the value of the deviation parameter $\Delta_g$ for this reflection  is directly depend on the value of temperature gradient.
 In the case of higher temperature of the back crystal face the neutron with  
$E_k - E_{k_g }>0$ will be reflected, while in the case of its lower temperature the neutron with  
$E_k - E_{k_g }<0$ will be reflected.  

\begin{figure}[ht]
\centerline{\psfig{file=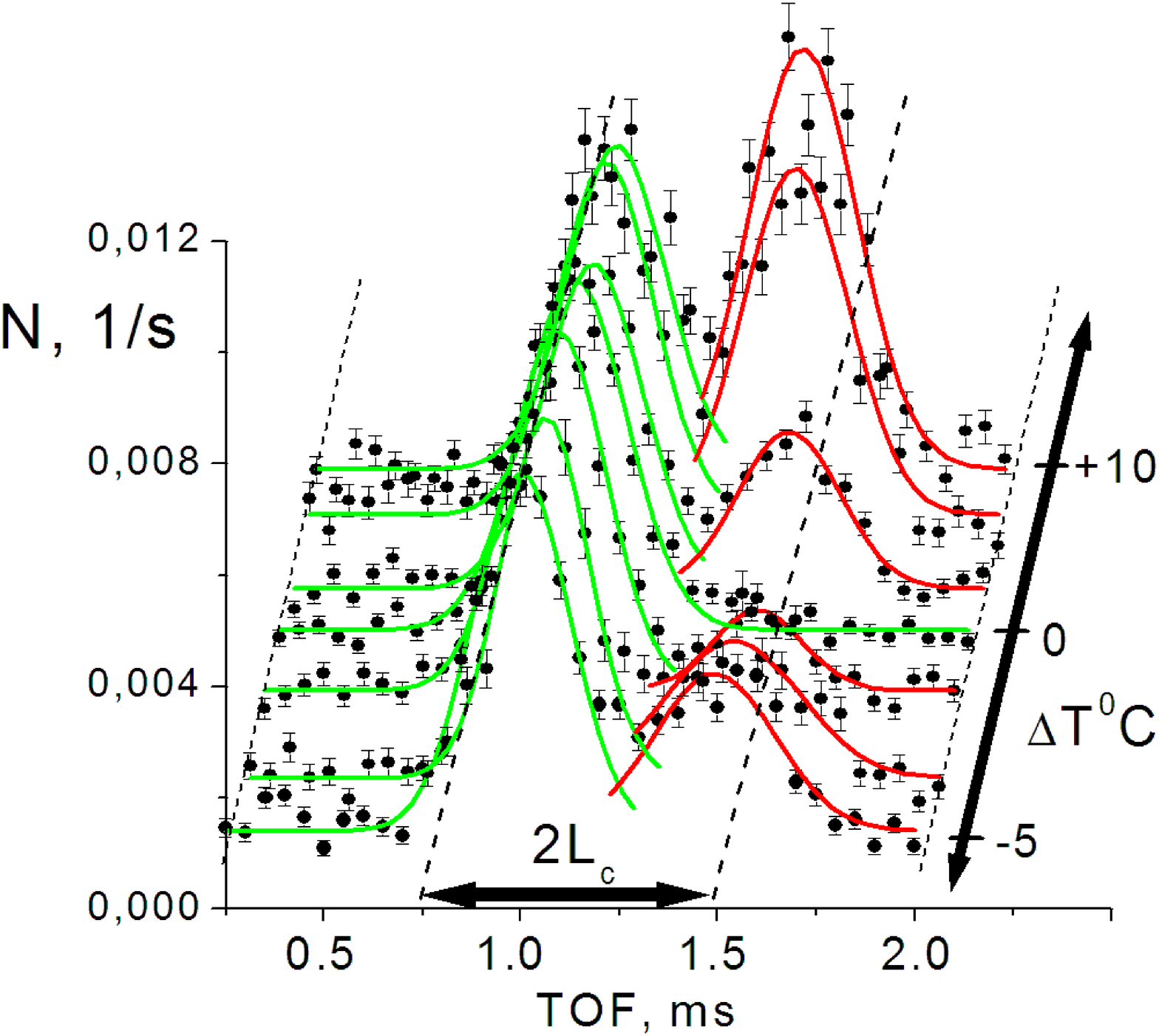,clip=,width=12 cm}}
\caption {Dependence of the time of flight (TOF) spectra of the neutron reflected by the (110) plane of quartz 
on the temperature gradient applied to the crystal. Bragg angle $\sim 90^0$, $L_c=27$ cm. One can see the reflexes from the front surface and from the back part of crystal.}
\label{fig:TOF_1}
\end{figure}

Examples of the time of flight spectra of the reflected neutrons for the Bragg angle
$\sim 90^0$ are shown in Fig.\ref{fig:TOF_1}. One can see  a formation of the reflex from the back crystal surface and increasing its intensity with the rise of the temperature gradient. 

The scheme of the experiment on the observation of neutron spin rotation is similar to that
described in \cite{PhysB2001}.

To observe the effect of neutron spin rotation due to Scwinger interaction it is necessary to turn the crystal in a position, for which Bragg angle is  different from $90^0$, because in the case of Bragg diffraction the Schwinger effect disappears for  $90^0$ Bragg angle:

\begin{equation}
\varphi _s = \frac{4 {\rm {\bf E}}  \mu  
L_c \mbox{v}_\parallel}{c\hbar \mbox{v}_ \bot } = \frac{4 {\rm {\bf E}} \mu L_c}{c\hbar 
}ctg(\theta _B )\mathrel{\mathop{\kern0pt\longrightarrow}\limits_{\theta _B 
\to \pi / 2}} 0
\end{equation}

The experiment on the observation of neutron spin rotation was carried out with the $L_c=14$ cm crystal thickness and Bragg angle $\approx 86^0$. 

\begin{figure}[htbp]
\centerline{\psfig{file=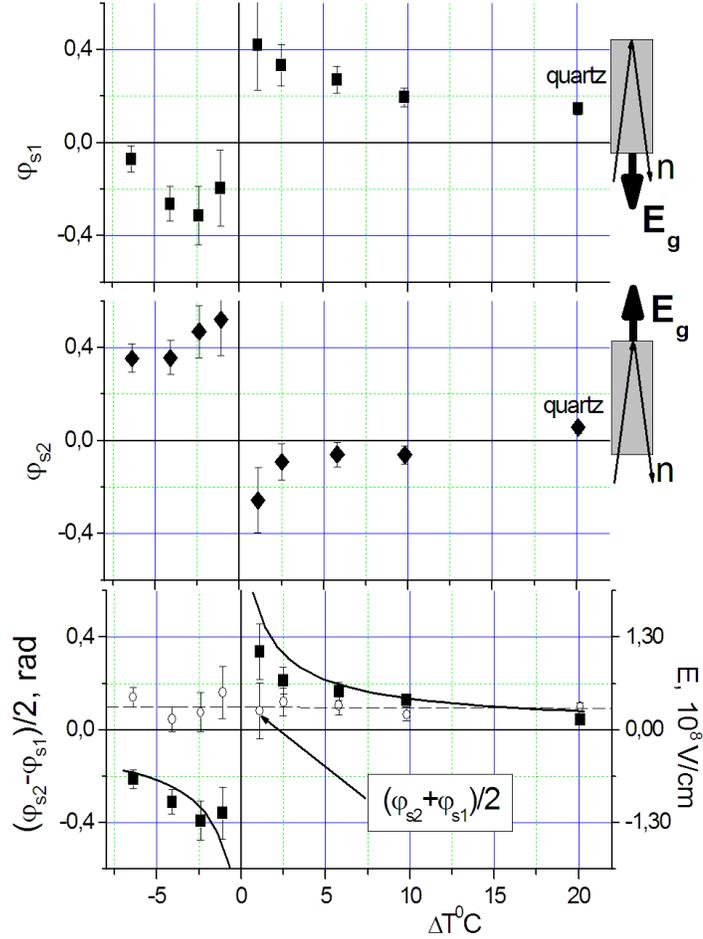,clip=,width=9.5 cm}}
\caption {The dependence of the angle of neutron spin rotation due to Schwinger interaction on the value of temperature gradient. Two upper figure corresponds to two crystal positions differing by the angle $180^0$. One can see a good coincidence of the theoretical dependence (solid curve in the bottom plot) with the experimental points.}
\label{fig:SR_1}
\end{figure}

The dependence of the angle of neutron spin rotation around ${\bf H}_S$ on the value of temperature gradient is shown in Fig.\ref{fig:SR_1}. 

We can change the sign of the effect by turn the crystal by the $180^0$ around ${\bf H}_S$. 
One can see that the experiment confirms that such a crystal rotation indeed change the sigh of the observed effect. On the right axis the effective electric field that is necessary to get the corresponding spin rotation effect is shown. One can see that the value of the electric field  reaches $\sim 1.3\cdot 10^8$ V/cm, that is only 1.5 times less than in the Laue diffraction case for exact Bragg condition \cite{dfield,dptfe}.

\section{Conclusion}
The first experimental study of some new phenomena for the neutron
diffraction and optics in  the
noncentrosymmetric crystal was carried out, using the pilot
set-up created for a search for the neutron EDM
by the crystal-diffraction me\-thod.

 For the first time  the neutron dynamical
Laue diffraction for the Bragg angles close to the right one
(up to $87^\circ$) was studied,
using the forward diffraction beam and the thick ($\sim$3.5 cm)
crystal.

The effect of the essential time delay of diffracting
neutrons inside the crystal for Bragg angles close to $90^\circ$
was experimentally observed.
For $(110)$-plane of $\alpha$-quartz and
$\theta_B=88.5^\circ$ we have obtained $\tau_L\approx 1.8$~ms that corresponds to $v_{\|}\approx 20$ m/s,
while $v=808$ m/s.

The predicted earlier phenomenon of the neutron beam depolarization
was first experimentally observed for the case of
Laue diffraction in the noncentrosymmetric $\alpha$-quartz
crystal.
It is experimentally proved that the interplanar electric field,
affecting the neutron in the crystal, maintains its value up to Bragg
angles equal to $87^\circ$.

It is shown experimentally that the value $E\tau$ determining
the sensitivity of the method in our case can reach $\sim 2\cdot
10^5$~V~s/cm that is comparable with that for the UCN method
($\sim 6\cdot 10^5$~V~s/cm)\cite{edmlast} and much more than the
value obtained by Shull and Nathans ($\sim 2\cdot
10^2$~V~s/cm)\cite{Shull} and also than that of \cite{Ramsey}
($\sim 1.2\cdot 10^3$~V~s/cm)\footnote{One can consider the
measurements of the depolarization as a first preliminary and
rough measurements of a neutron EDM, which gives the result $D <
10^{-22}$ e$\cdot$cm that is some better than the old Shull's
and Nathans's result  \cite{Shull}. The result obtained in
magnetic resonance method using cold neutrons \cite{Ramsey} was
$D < 3\cdot 10^{-24}$ e$\cdot$cm.}.
 
These results give the opportunity to try the Laue diffraction method for a
neutron EDM search. 
The statistical sensitivity of the  method  was estimated experimentally to be  
$\sigma(D)\approx 6\times 10^{-25}~e\cdot{\rm cm}$ per day for the PF1B beam of the
ILL reactor and available quartz crystal. The measured intensities
and the neutron time of stay in the quartz crystal coincide well with the
theoretical predictions.
The use of
the other crystals may allow to improve the sensitivity of the method by about one
order of magnitude  and to reach $\sim$ a few $10^{-26}$e$\cdot$~cm per day.

The effect of a spin rotation due to Schwinger interaction
of the magnetic moment of moving neutron with an interplanar
electric field of the noncentrosymmetric crystal was
experimentally observed for  neutron, passing through the
crystal far from the Bragg conditions.

The energy dependence of a spin rotation angle was measured for two
crystal orientations. The dependence  has a pronounced
resonance form. The direction of a spin rotation changes its sign when the energy  passes  through the Bragg resonance value. So a new kind of neutronography arises for noncentrosymmetric crystals, which allows to "see"\ and study the crystallographic planes with nonzero electric interplanar fields. 

For $\alpha$-quartz crystal the value of the spin rotation angle
can reach $\pm (1 - 2)\cdot 10^{-4}$~rad/cm that corresponds to the value of
resultant electric field equal to $\pm
(0.5 - 1)\cdot 10^{5}$~V/cm.

We note also that the presence of the  terms (\ref{v'}) and (\ref{E'})
in equation (\ref{Etild}) should result in the dependence of a
neutron absorption  on the direction and the value of
a neutron wave vector as well as on the
neutron spin orientation.

The effect of spin rotation in a noncentrosymmetric quartz crystal for neutrons Bragg reflected by the deformed part of crystal was first observed.
This effect is caused by the Schwinger interaction and depends on a deformation degree of crystal near its back surface. For the quartz crystal the effective electric field affected the neutron during the time of its staying inside the crystal can reach  $\sim 1.3\cdot 10^8$~V/cm.
Simple estimation has shown that in our case the depth of neutron penetration into the crystal and so the time of neutron interaction with the electric field can be about four or even five orders more than in the well known Shull and Nathans experiment for the neutron EDM search \cite{Shull}.

In addition, the requirements to the crystal perfection are relatively low for this scheme. For the case $\gamma_B << w_m$ the effective electric field affected the neutron depends on an effective crystal mosaicity $w_m$ as $E=E_0(\gamma_B/w_m)$, where $\gamma_B$ is the angular Bragg width, but the reflex intensity increase as $I=I_0(w_m/\gamma_B)$, therefore the sensitivity to measure the neutron EDM will be reduced only by a factor $\sqrt{w_m/\gamma_B}$, that give us a hope that such a scheme can be applied to search the T-odd part of neutron-nuclei interaction \cite{bar} using neutrons with energies near the P wave resonance one.

The authors are grateful to all our colleagues
E.G. Lapin, E. Lelie`vre-Berna, V. Nesvizhevsky, A. Petoukhov, S.Yu. Semenikhin, T. Soldner, F. Tasset for numerous and
useful discussions and for the active participation in the experiments.

This work is supported by RFBR: grants N~03-02-17016, 05-02-16241,
by INTAS: grant N~00-00043.


\begin{thebibliography}{10}

\bibitem{Khriplovich}  Khriplovich,I.B.;  Lamoreaux,S.K.
{\it CP Violation without Strangeness. The Electric Dipole Moments of
Particles, Atoms and Molecules}; Springer-Verlag; 1996.

\bibitem{Bunakov}  Bunakov,V.E. Fiz. Elem. Chast. Atom. Yad. (EChAYa) 1995 {\bf 26}
285--361. (in russian)

\bibitem{pnpiedm} Altarev,I.S.; Borisov,Yu.V.; Borovikova,N.V.;
et al. Yad.Fiz. 1996 {\bf 59} 1204.

\bibitem{illedm} Smith,K.F.; Crampin,N.; Pendlebury,J.M.; et
al. Phys. Lett. 1990 {\bf B234} 191.

\bibitem{edmlast} Harris,P.G.; Baker,C.A.; Green,K.; et al.
Phys. Rev. Lett. 1999 {\bf 82} 904.

\bibitem{NIMB_FedVor} Fedorov,V.V.; Voronin,V.V.
Nucl. Instr. and Meth. B 2003 {\bf B201} (1)  230. 

\bibitem{dfield} Alexeev,V.L.; Fedorov,V.V.; Lapin,E.G.; Leushkin,E.K.;
      Rumian\-t\-sev,V.L.; Sumbaev,O.I.; Voronin,V.V.
Nucl. Instr. and Meth. A 1989 {\bf A284}  181;\\
Sov. Phys. JETP 1989 {\bf 69}  1083.

\bibitem{dedm} Fedorov,V.V.; Voronin,V.V.; Lapin,E.G. Preprint
LNPI-1644, Leningrad 1990 36p.;\\J. Phys. G 1992 {\bf 18} 1133.

\bibitem{polart} Fedorov,V.V.; Voronin,V.V.; Lapin,E.G.;
Sumbaev,O.I. Preprint PNPI-1944, Gatchina 1994  10p.;\\ Tech.
Phys. Lett. 1995 {\bf 21}(11) 881;\\ Physica B 1997 {\bf 234--236} 8.

\bibitem{Forte} Forte,M. J. Phys. G 1983 {\bf 9} 745.

\bibitem{Shull} Shull,C.G.; Nathans,R. Phys. Rev. Lett.
1967 {\bf 19} 384.

\bibitem{Barysh}  Baryshevskii,V.G.; Cherepitsa,S.V.
Phys. Stat. Sol. 1985 {\bf b128} 379;\\
Izvestiya Vuzov SSSR, ser. fiz. 1985 {\bf 8 } 110 (in Russian).

\bibitem{GolPendl} Golub,R.; Pendlebury,G.M.
Contemp. Phys.  1972 {\bf 13}  519.

\bibitem{3} Abov,Yu.G.;  Gulko,A.D.; Krupchitsky,P.A. {\it Polarized Slow
    Neutrons}; Atomizdat; Moscow, 1966;  256p. (in russian).

\bibitem{sdprepr}
 Alexeev,V.L.;  Voronin,V.V.;  Lapin,E.G.;  Leushkin,E.K.;
      Rumiantsev,V.L.;   Fedorov,V.V.
Preprint LNPI--1608, Leningrad 1990 12p.;\\
Tech. Phys. Lett. 1995 {\bf 21}(11) 881.

\bibitem{fedvor} Fedorov,V.V.; Voronin,V.V. in
Physics of Atomic Nuclei and Elementary Particles,
Proc. of the XXX \,PNPI Winter School;
 St.Petersburg, 1996; 123 (in russian).

\bibitem{Rauch1} Schuster,M.; Carlile,C.; Rauch,H. Z.Phys.
1991 {\bf B85} 49;\\
Jericha,E.; Carlile,C.;  Rauch,H. Nucl. Inst. and Meth. 1996 {\bf
A379} 330.

\bibitem{Domb} Dombeck,T. ANL-report PHY-8624-HI-97;\\
Dombeck,T.; Kaiser,H.; Koetke,D.; Peshkin,M.;
Ringo,R. ANL-report PHY-9814-TH-2001;\\
Dombeck,T.; Ringo,R.; Koetke,D.D.; Kaiser,H.;
Schoen,K.; Werner,S.A.; Dombeck,D. Phys. Rev. A 2001 {\bf 64} 053607.

\bibitem{ForteZeyen} Forte,M.; Zeyen,C.M.E.
Nucl. Instr. and Meth. A 1989 \textbf{A284} 147.

\bibitem{PbTiO3}
Voronin,V.V; Fedorov,V.V.; Preprint PNPI-2293, Gatchina 1999 10p.

\bibitem{bar}
Baryshevsky,V.G.  J.Phys. G 1997 \textbf{23} 509.

\bibitem{Hirsh} Hirsch,P.B.; Howie,A.; Ni\-chol\-son,R.B.; et
al., Electron Mic\-ros\-co\-py of Thin
Crystals; Butterworths; London, 1965.

\bibitem{Kirian} Fedorov,V.V.; Kir'yanov,K.E.; Smirnov,A.I.
Sov. Phys. JETP 1973 {\bf 37} 737.

\bibitem{RauchPetr} Rauch,H.; Petrachek,D.
in Neutron diffraction; Ed. Duchs,H.; Dynamical neutron diffraction and its application;
Springer, Berlin, 1978; 303.

\bibitem{Golub}  Golub,R.;  Lamoureux,K.L., Phys. Rep. 1994 {\bf 237} 1.

\bibitem{PhysB2003} Voronin,V.V.; Fedorov,V.V.; Lapin,E.G.; 
Semenikhin,S.Yu. Physica B 2003 \textbf{335} (1-4) 201-204.

\bibitem{Appl_Ph_DEDM} Fedorov,V.V.; Lapin,E.G.; Semenikhin,S.Yu.; Voronin,V.V. 
Appl. Phys. A 2002 \textbf{74}[Suppl.] s91-s93.

\bibitem{tfjetpl}
Voronin,V.V.;  Lapin,E.G.; Semenikhin,S.Yu.; Fedorov,V.V.
JETP Lett. 2000 {\bf 71}(2) 76.
\\ Preprint PNPI-2337; Gatchina 2000  12p.

\bibitem{Nopt_NIMB}  Fedorov,V.V.; Voronin,V.V. 
Nucl. Instr. and Meth. B 2003 \textbf{201} 230-242.

\bibitem{dptfe}
      Voronin,V.V.;  Lapin,E.G.; Semenikhin,S.Yu.; Fedorov,V.V.
      Preprint PNPI-2376, Gatchina 2000  15p.;
 JETP Lett. 2000 {\bf 72}(6) 308.

\bibitem{LDM_sens} Fedorov,V.V.; Lapin,E.G.; Lelie`vre-Berna,E.; Nesvizhevsky,V.; Petoukhov,A.; Semenikhin,S.Yu.; Soldner,T.; Tasset,F.; Voronin,V.V.
Nuclear Inst. and Methods in Physics Research B 2005 \textbf{227} (1-2) 11-15.

\bibitem{PhysB2001} Fedorov,V.V.; Lapin,E.G.; 
Semenikhin,S.Yu.; Voronin,V.V. Physica B 2001 {\bf
297}(1-4) 293.

\bibitem{dedm2}
Fedorov,V.V. in Proc. of XXVI Winter LNPI School; Leningrad, 1991; Vol.1,
65.

\bibitem{Appl_Ph_Srot} Fedorov,V.V.; Lapin,E.G.; Semenikhin,S.Yu.; Voronin,V.V. 
Appl. Phys. A 2002 \textbf{74}[Suppl.] s298-s301.

\bibitem{srjetpl}
Voronin,V.V.;  Lapin,E.G.; Semenikhin,S.Yu.; Fedorov,V.V.
JETP Lett. 2001 {\bf 74}(5) 251-254.
\\ Preprint PNPI-2431; Gatchina 2001  14p.

\bibitem{Ramsey} Dress,W.B.; Miller,P.D.; Pendlebury,J.M.; Perrin,P.;
Ramsey,N.F. Phys. Rev. D 1977 {\bf 15}(1) 9.


\end{thebibliography}
\end{document}